\documentclass[prl,reprint, twocolumn,showpacs,preprintnumbers,amsmath,amssymb,nofootinbib,floatfix]{revtex4-1} 
\usepackage{graphicx}

\usepackage{mathrsfs}
\usepackage{hyperref}

\usepackage{slashed}

\usepackage{color}

\usepackage{gensymb}

\def \matrix #1 {\left(\begin{array}{cc} #1 \end{array}\right)}

\def\II{\hbox{{1}\kern-.25em\hbox{l}}}

\begin{document}

\title{Deciphering the long-distance penguin contribution to $\bar B_{d, s} \to \gamma \gamma$ decays }

\author{Qin Qin$^{a}$}
\email{qqin@hust.edu.cn}

\author{Yue-Long Shen$^{b}$}
\email{corresponding author: shenylmeteor@ouc.edu.cn}

\author{Chao Wang$^{c}$}
\email{corresponding author: chaowang@nankai.edu.cn}

\author{Yu-Ming Wang$^{d}$}
\email{corresponding author: wangyuming@nankai.edu.cn}

\affiliation{${}^a$ School of Physics, Huazhong University of Science and Technology,
Luoyu Road 1037, Wuhan Hubei 430074, P.R. China  \\
${}^b$ College of Information Science and Engineering, Ocean University of China,
Songling Road 238, Qingdao, Shandong 266100, P.R. China \\
${}^c$ Department of Mathematics and Physics,
Huaiyin Institute of Technology,
Meicheng East Road 1,  Huaian, Jiangsu 223200, P.R. China \\
${}^d$ School of Physics, Nankai University,
Weijin Road 94, Tianjin 300071, P.R. China}

\date{\today}

\begin{abstract}
\noindent
We compute for the first time the long-distance penguin contribution to the double radiative $B$-meson decays
due to the purely hadronic operators acting with the electromagnetic current in the background soft-gluon field
from first field-theoretical principles by introducing a novel subleading $B$-meson distribution amplitude.
The  numerically dominant penguin amplitude arises from the soft-gluon radiation off
the light up-quark loop rather than the counterpart charm-loop effect on account of
the peculiar analytical behaviour of the short-distance hard-collinear function.
Importantly the  long-distance up-quark penguin contribution brings about the substantial cancellation
of the known factorizable power correction possessing the same multiplication CKM parameters,
thus enabling $B_{d, \, s} \to \gamma \gamma$ to become new benchmark probes of  physics beyond the Standard Model.
\\[0.4em]

\end{abstract}


\maketitle

%
\section{Introduction}
%

It is widely accepted that the exclusive radiative penguin bottom-meson decays play a central role
in exploring the quark-flavour dynamics of the Standard Model (SM)
and in probing the nonstandard electroweak interactions at the LHCb and Belle II experiments.
In particular, the double radiative $\bar B_{d, s} \to \gamma \gamma$ decays with non-hadronic final states
offer a remarkably clean environment to address the intricate strong interaction mechanism
of the heavy-hadron system with the perturbative factorization technique,
in comparison with the  radiative  decays $\bar B \to V \gamma$.
Phenomenologically the direct CP asymmetries of the double radiative $B$-meson decays
with the linearly polarized photon states will be also highly beneficial
for determining the CKM phase angle $\gamma$ \cite{Bosch:2002bv}.
Applying the QCD factorization approach the leading-power contributions to
the exclusive $\bar B_{d, s} \to \gamma \gamma$ decay amplitudes in the heavy quark expansion
have been demonstrated to be factorized into the short-distance Wilson coefficients due to the hard
and hard-collinear fluctuations as well as
the leading-twist bottom-meson distribution amplitude \cite{Descotes-Genon:2002lal}.
In addition, a variety of the subleading-power corrections to the radiative penguin
$\bar B_{d, s} \to \gamma \gamma$ decay amplitudes (including the higher-twist off light-cone correction
and the non-leading Fock-state effect)  were  investigated
at tree level in the strong coupling constant with the diagrammatic factorization approach \cite{Shen:2020hfq}.

However, the persistent problem of evaluating the long-distance penguin contribution
to the double radiative bottom-meson decay amplitudes in the presence of soft gluon emission
remains unresolved at present.
For decades the non-local subleading power correction arising from
the soft gluon radiation off the charm-loop diagrams  constitutes the longstanding obstacle
to improve theory computations for the  angular observables of $B \to K^{(\ast)} \ell \ell$
at large hadronic recoil \cite{Ali:1991is,Khodjamirian:2010vf,Khodjamirian:2012rm,Hambrock:2015wka,Bobeth:2017vxj,Kozachuk:2018yxf,Melikhov:2019esw,Gubernari:2020eft}
(see \cite{Grinstein:2004vb,Beylich:2011aq,Lyon:2014hpa} for more discussions on such non-local contribution at low hadronic recoil).
Achieving the robust predictions of the long-distance charm-loop effect in the rare $B \to K^{(\ast)} \ell \ell$ decays
will be evidently indispensable for disentangling the genuine New Physics (NP) effect from the SM background contribution
and for advancing our understanding towards the nature of the observed flavour anomalies
(see for instance \cite{Jager:2012uw,Descotes-Genon:2013wba,Descotes-Genon:2015uva,Ciuchini:2015qxb,Aebischer:2019mlg,Ciuchini:2020gvn}).
To this end, constructing the systematic  theory formalism to tackle the long-distance penguin contribution
to $\bar B_{d, s} \to \gamma \gamma$ will further shed new light on the model-independent calculation of
the analogous QCD corrections to the exclusive flavour-changing neutral current (FCNC) decays
$B \to K^{(\ast)} \ell \bar \ell$.
More generally, the newly proposed framework to cope with the non-local  power correction to the double radiative
$\bar B_{d, s}$-meson  decays  will be in the meanwhile  of paramount importance to perform the precision calculation of
the radiative and electroweak penguin decays of heavy-flavour baryons \cite{Chen:2001zc,He:2006ud,Wang:2008sm,Ball:2008fw,Wang:2009hra,Mannel:2011xg,Feldmann:2011xf,Wang:2011uv,Braun:2014npa,Wang:2015ndk}.

According to the numerical hierarchy between the bottom and charm quark masses,
we will apply the favored  power counting scheme $m_b \gg m_c \sim {\cal O} (\sqrt{\Lambda \, m_b}) \gg \Lambda$
as advocated in \cite{Grinstein:2004vb,Boos:2005by,Boos:2005qx,Wang:2017jow,Gao:2021sav}
in establishing the perturbative factorization formulae
for the penguin contractions of the effective four-quark operators accompanied by the soft gluon emission,
instead of the alternative counting scheme $m_b \sim m_c \gg \Lambda$ implemented in the exclusive two-body
$B$-meson decays \cite{Beneke:2000ry,Huber:2016xod}.
Subsequently, we will report on a novel observation on the very hadronic matrix element
responsible for the soft gluon radiation off the penguin diagrams.
Integrating out the short-distance QCD fluctuations  embedded in this hadronic quantity
will give rise to the generalized three-particle $B$-meson distribution amplitudes in heavy quark effective theory (HQET)
defined by the non-local matrix element
$\langle 0 | \bar q_{s}(\tau_1 n) \, G_{\mu \nu}(\tau_2 \bar n) \, \Gamma_i \,  h_v(0) | \bar B_v\rangle$
(in analogy to the subleading shape function $g_{17}(\omega, \omega_1, \mu)$ discussed in \cite{Benzke:2010js})
rather than the conventional light-cone distribution amplitudes as previously introduced in \cite{Kawamura:2001jm,Braun:2017liq}.
Employing the asymptotic behaviour of the generalized $B$-meson distribution amplitudes
at small quark and gluon momenta and the model-independent theory constraints on these non-perturbative functions,
we will proceed to demonstrate that the soft-collinear convolution integrals entering the factorized expressions
of the long-distance penguin contributions converge for both the massless-quark and massive-quark loop induced pieces,
by contrast with the corresponding mechanism in the FCNC decay processes $B \to K^{(\ast)} \ell \ell$.
Phenomenological implications of the newly computed power correction to the double radiative bottom-meson decay observables
will be further explored with the aid of the three-parameter model for the particular subleading  $B$-meson
distribution amplitude of our interest.

%

%
\section{General analysis}
%

The effective weak Hamiltonian of the double radiative $b \to q \gamma \gamma$
transitions has been shown to be identical to the one for  $b \to q \gamma$ decays \cite{Grinstein:1990tj}
\begin{eqnarray}
{\cal H}_{\rm eff} &=&  {4 \, G_F \over \sqrt{2}} \, \sum_{p=u, c} \, V_{p b} V_{p q}^{\ast}  \,
\bigg [ C_1(\nu) \, P_1^{(p)}(\nu) + C_2(\nu) \, P_2^{(p)}(\nu)
\nonumber \\
&& + \sum_{i=3}^{8} C_i(\nu) \, P_i(\nu) \bigg ] + {\rm h.c.}   \,,
\label{effective weak Hamiltonian of b to gamma gamma}
\end{eqnarray}
by employing the classical equations of motion \cite{Politzer:1980me}.
We will further adopt the effective operator basis   $P_i^{(p)}$  as advocated in \cite{Chetyrkin:1996vx}
ensuring  the disappearance of Dirac traces involving an odd number
of $\gamma_5$ in the subsequent effective theory computations.

Up to the lowest order in the electromagnetic interaction
one can conventionally cast the exclusive radiative decay amplitude
for $\bar B_q \to \gamma \gamma$ in the following form \cite{Shen:2020hfq}
\begin{eqnarray}
{\cal \bar A}(\bar B_q \to \gamma \gamma)
 &=& - {4 \, G_F \over \sqrt{2}} \, {\alpha_{\rm em} \over 4 \pi}  \,
\epsilon^{\ast \alpha}(p)  \, \epsilon^{\ast \beta}(q)  \,
\nonumber \\
&& \times \, \sum_{p=u, c} \, V_{p b} V_{p q}^{\ast}  \,
\sum_{i=1}^{8} C_i \, T_{i, \, \alpha \beta}^{(p)} \,,
\label{general parametrization for the radiative decay amplitude}
\end{eqnarray}
where the yielding  hadronic tensors $T_{i, \, \alpha \beta}^{(p)}$
can be decomposed into the helicity form factors
\begin{eqnarray}
T_{i, \, \alpha \beta}^{(p)}  &=&
i \, m_{B_q}^3 \, \bigg  [ \left (g_{\alpha \beta}^{\perp} - i \, \varepsilon_{\alpha \beta}^{\perp} \right ) \, F_{i, L}^{ (p)}
 -  \left (g_{\alpha \beta}^{\perp} + i \, \varepsilon_{\alpha \beta}^{\perp} \right ) \, F_{i, R}^{ (p)}  \bigg  ],
\nonumber \\
\end{eqnarray}
thanks to the QED Ward-Takahashi identities and the transversality of the on-shell photons.
Here we have introduced the shorthand notations for brevity
\begin{eqnarray}
g_{\alpha \beta}^{\perp} \equiv g_{\alpha \beta}-{n_{\alpha}  \bar n_{\beta} \over 2}
-{\bar n_{\alpha}   n_{\beta} \over 2},  \hspace{0.3 cm}
\varepsilon_{\alpha \beta}^{\perp} \equiv {1 \over 2} \, \varepsilon_{\alpha \beta \rho \tau} \bar n^{\rho} n^{\tau},
\hspace{0.2 cm}
\end{eqnarray}
by defining  two light-cone vectors $n_{\mu}$ and $\bar n_{\mu}$
which satisfy the kinematic constraints $p_{\mu} = m_{B_q} \, \bar n_{\mu}/2$
and $q_{\mu} = m_{B_q} \, n_{\mu}/2$.
It is  interesting to note that  only the left-handed  form factors $F_{i, L}^{ (p)}$
will survive  at leading order in the heavy quark expansion on account of the helicity conservation
of the QCD interaction at high energy.
Explicitly, the resulting factorization formula for  $F_{i, L}^{ (p)}$ at leading power can then be written as
\begin{eqnarray}
&& \sum_{i=1}^{8} C_i \, F_{i, L}^{(p), \,  {\rm LP}} =
-  {Q_q \, f_{B_q} \, \overline{m}_b(\nu)   \over m_{B_q}} \,
\, V_{7, \, \rm{eff}}^{(p)}(m_b, \mu, \nu)  \,
\nonumber \\
&&   \hspace{1.2 cm}  \times \,  K^{-1}(m_b, \mu) \,  \int_0^{\infty} \, { d \omega \over \omega} \, \phi_{B}^{+}(\omega, \mu) \,
J (m_{b}, \omega, \mu),
\hspace{0.8 cm}
\end{eqnarray}
where the desired expressions for the effective hard function $V_{7, \, \rm{eff}}^{(p)}$, the perturbative matching coefficient $K$
and the hard-collinear function $J$ at the one-loop accuracy can be found in \cite{Shen:2020hfq,Eichten:1989zv,Lunghi:2002ju,Bosch:2003fc}.

%
\section{QCD factorization for  the long-distance penguin contribution}
%

\begin{figure}[tp]
\begin{center}
\includegraphics[width=0.75 \columnwidth]{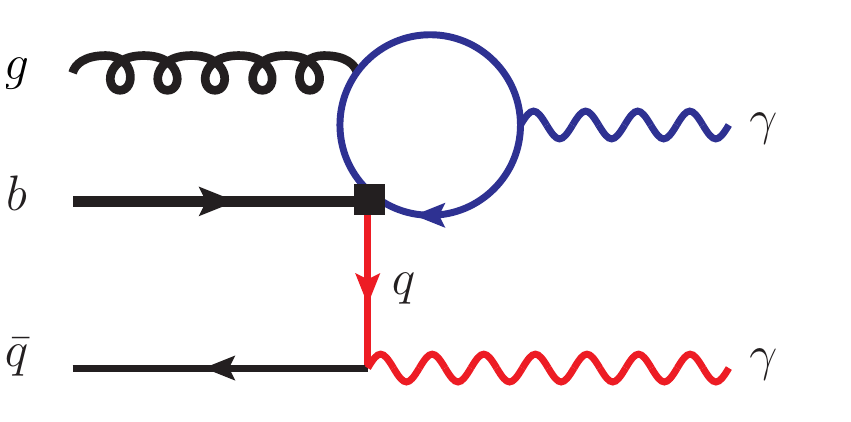}
\caption{Diagrammatical representation of the soft-gluon emission from
the factorizable quark loop  generated by the effective four-quark operator
and the electromagnetic current  in the double radiative $\bar B_q \to \gamma \gamma$ decay process,
where the symmetric diagram due to the exchange of
two on-shell photons is not presented.}
\label{fig: Feynman diagrams for the nonfactorizable quark loops}
\end{center}
\end{figure}

We are now in a position to explore factorization properties  of  the long-distance penguin contribution
to the double radiative $\bar B_q \to \gamma \gamma$ decay amplitude
by inspecting the partonic diagram displayed in Figure \ref{fig: Feynman diagrams for the nonfactorizable quark loops}.
Integrating out the hard-collinear quark loop one can readily derive
the flavour-changing scattering amplitude of $g (\ell)+ b (v)\to q (\tilde{q}) + \gamma(p)$
governed by the effective weak Hamiltonian (\ref{effective weak Hamiltonian of b to gamma gamma})
by discarding the subleading-power terms  in $\Lambda/m_b$ and by invoking the on-shell constraint
for the external photon state
\begin{eqnarray}
&& {\cal M} (g + b \to q + \gamma)  =  i \, {4 \, G_F \over \sqrt{2}} \,  {g_{\rm em} g_s \over 4 \pi^2}\,
\sum_{p=u, c} \, V_{p b} V_{p q}^{\ast} \,
\bigg \{
\nonumber \\
&&  \left (C_2 - {C_1 \over 2 N_c}  \right)  \, Q_p \, \left [ F(z_p) - 1 \right ]
+ 6 \, C_6 \, \sum_{q^{\prime}} \, Q_{q^{\prime}} \, \left [ F(z_{q^{\prime}}) - 1 \right ]
\nonumber \\
&& + \left[  \left (C_3 - {C_4 \over 2 N_c}  \right) + 16 \,
\left (C_5 - {C_6 \over 2 N_c}  \right)   \right ] \,
\, Q_{q} \, \left [ F(z_{q}) - 1 \right ]
\bigg \}
\nonumber \\
\nonumber \\
&&   \times \, \left [ \bar q (\tilde{q}) \,  \gamma_{\beta} \, P_L \, G_{\mu \alpha} \,\,
\tilde{F}^{\mu \beta}b(v) \right ] \, \frac{p^{\alpha}} {(p-\ell)^2} \,,
\label{scattering amplitude of g b to q gamma}
\end{eqnarray}
where the perturbative penguin function is given by
\begin{eqnarray}
F(x)= 4 \, x \, \arctan^{2} \left ( \frac{1}{\sqrt{4 \, x -1}}  \right ) \,,
\label{penguin function}
\end{eqnarray}
and we have further employed  the conventions
\begin{eqnarray}
z_p = {m_p^2 - i \, 0^{+} \over (p-\ell)^2} \,,
\qquad
\tilde{F}^{\mu \nu} = - \left ( {1 \over 2} \right ) \, \epsilon^{\mu \nu \alpha \beta} \, F_{\alpha \beta}\,.
\end{eqnarray}
Apparently, the hard-scattering kernel of the partonic amplitude displayed in
(\ref{scattering amplitude of g b to q gamma}) depends on the unique component
$\bar n \cdot \ell$  of the soft-gluon momentum at leading order in the heavy quark expansion.
Moreover, it is straightforward to verify that the bottom-quark loop diagram with an insertion of the QCD penguin operator $P_6$
can only bring about the subleading-power effect, by virtue of the asymptotic behaviour of
$F(x) - 1 \sim  {\cal O} ({1 / x})$ at large $x$,  as it should be.
It remains necessary to remark that the long-distance penguin contraction mechanism does not give rise to
the non-trivial strong phase for the radiative $\bar B_q \to \gamma \gamma$ amplitude in consequence of
the space-like four-momentum $(p - \ell)$,
by contrast with the counterpart contribution to the radiative leptonic $\bar B_q \to \gamma \ell \bar \ell$  decay.

We can proceed to evaluate the five-point amplitude of $g (\ell)+ b (v) + \bar q (k) \to \gamma(p) + \gamma(q)$
with the diagrammatic factorization technique by integrating out the anti-hard-collinear quark propagator
in Figure \ref{fig: Feynman diagrams for the nonfactorizable quark loops} subsequently
\begin{eqnarray}
&& \langle \gamma(p) \,  \gamma(q) | \bar q \,  \gamma_{\beta} \, P_L \, G_{\mu \alpha} \,\,
\tilde{F}^{\mu \beta} \, b  | g (\ell) \,  b (v)  \, \bar q (k) \rangle
\nonumber  \\
&& \Rightarrow {i \, g_{\rm em} \, e_q \over (q-k)^2} \, \epsilon^{\mu \beta \lambda \tau} \, p_{\lambda} \,
\epsilon_{\tau}^{\ast}(p)  \, \epsilon_{\rho}^{\ast}(q)  \,
\nonumber \\
&& \hspace{0.4 cm}
\times \, \left [ \bar q(k) \, \gamma_{\perp}^{\rho}  \not \! {q} \, \gamma_{\beta} \, P_L \, G_{\mu \alpha}(\ell)  \, b(v) \right ] \,
+ {\cal O}(\alpha_s) \,,
\end{eqnarray}
where the yielding short-distance matching coefficient  depends on the component $n \cdot k$
(rather than  $\bar n \cdot k$) of the soft-quark momentum in the leading-power approximation.
As a consequence, it becomes evident to introduce the subleading $B$-meson distribution amplitude
defined by  the HQET matrix element of the  three-body non-local operator
$\bar q_{s}(\tau_1 n) \, G_{\mu \nu}(\tau_2 \bar n) \, \Gamma_i \,  h_v(0)$
for the sake of describing  the soft QCD dynamics encoded in  the long-distance penguin contribution
to $\bar B_q \to \gamma \gamma$. Constructing the general parametrization of the emerged  effective matrix element
with the covariant tensor formalism \cite{Falk:1990yz}
(in analogy to the Lorentz decomposition for the corresponding light-cone matrix element \cite{Braun:2017liq})
allows us to derive the soft-collinear factorization formula of the soft-gluon radiative correction
to the left-handed helicity form factor
\begin{eqnarray}
&& \sum_{i=1}^{8} C_i \, F_{i, L}^{(p), \,  {\rm soft \, 4q}} =
-  {Q_q \, f_{B_q} \over m_{B_q}} \, \int_0^{\infty} {d \omega_1 \over  \omega_1}\,
\int_0^{\infty} {d \omega_2 \over  \omega_2} \,
\nonumber \\
&& \,\,  \bigg \{  \left (C_2 - {C_1 \over 2 N_c}  \right)  \, Q_p \, \left [ F(z_p) - 1 \right ]
+ 6 \, C_6 \, Q_c \, \, \left [ F(z_c) - 1 \right ]
\nonumber \\
&& \,\,  - \left[  \left (C_3 - {C_4 \over 2 N_c}  \right) + 16 \,
\left (C_5 - {C_6 \over 2 N_c}  \right)   \right ] \,
\, Q_{q}  \bigg \} \,
\nonumber \\
\nonumber \\
&& \,\,  \times  \,  \Phi_{\rm G}(\omega_1, \omega_2, \mu)
+ \, {\cal O}(\alpha_s)\,,
\hspace{0.5 cm}
\label{LO factorization formula}
\end{eqnarray}
where the electric-charge relation for the light-flavour quarks
$Q_u + Q_d + Q_s =0$ and the vanishing penguin function $F(0)$
have been applied to simplify the obtained perturbative kernel.
It is customary to define two light-cone variables $\omega_{1}= n \cdot k$ and $\omega_{2}=\bar n \cdot \ell$
such that the resulting hard-collinear function develops a peculiar dependence on  $\omega_2$
via the dimensionless quantity  $z_p= - m_p^2 /(m_B \, \omega_2)$
(apart from an overall factor $1 / \omega_2$).
Including the higher-order QCD corrections to the partonic diagrams
in Figure \ref{fig: Feynman diagrams for the nonfactorizable quark loops}
will generate the non-trivial hard functions (instead of ``$1$" at tree level)
from matching the effective four-quark operators $P_i^{(p)}$ onto ${\rm SCET}_{\rm I}$
and simultaneously  result in the interesting impacts on the (anti)-hard-collinear matching coefficients
appeared in (\ref{LO factorization formula}).
However, the fundamental property that the determined short-distance matching functions
merely depend on the two dimensional  variables $\omega_1$  and $\omega_2$ remains valid
beyond the leading-order accuracy, thus justifying the appearance  of the soft matrix element
$\langle 0 | \bar q_{s}(\tau_1 n) \, G_{\mu \nu}(\tau_2 \bar n) \, \Gamma_i \,  h_v(0) | \bar B_v\rangle$
(see \cite{Beneke:2020fot} for further discussions in a different context).
Schematically, the factorized expression for the long-distance penguin correction to $\bar B_{q} \to \gamma \gamma$
can be cast in the form ${\cal H}  \, {\cal J} \star \bar {\cal J} \star \Phi_{\rm G}$,
which resembles the very pattern for the $Q_1^{q}-Q_{7 \gamma}$ contribution to $\bar B \to X_s \gamma$ \cite{Benzke:2010js}.
We also mention in passing that the newly computed long-distance soft gluon radiative correction
appears to preserve the large-recoil symmetry of the two transversality amplitudes
and does not affect the right-handed form factors $F_{i, R}^{ (p)}$, in agreement with the earlier observation
on the ``resolved photon" contribution to $\bar B \to \gamma \ell \nu$ at twist-two \cite{Wang:2018wfj}
(see also \cite{Braun:2012kp,Wang:2016qii} for an estimate of the soft contribution with the dispersion approach).

Additionally, the novel subleading distribution amplitude (perhaps more appropriately called {\it soft function})
of the $B$-meson $\Phi_{\rm G}$ in  (\ref{LO factorization formula})
is defined in terms of the effective  matrix element of the three-body non-local operator
with quark and gluon fields localized on two distinct light-cone directions
\begin{widetext}
\begin{eqnarray}
&& \langle 0 | (\bar q_{s} S_{n}) (\tau_1 n) \, (S_{n}^{\dagger} \, S_{\bar n})(0) \,
(S_{\bar n}^{\dagger} \, g_s \, G_{\mu \nu} \, S_{\bar n} )(\tau_2 \bar n) \,\,
\bar n^{\nu} \not \! n \gamma_{\perp}^{\mu} \gamma_5 \,
(S_{\bar n}^{\dagger} h_v) (0) | \bar B_v \rangle
\nonumber \\
&& = 2 \, \tilde{f}_B(\mu) \, m_B \, \int_0^{\infty} d \omega_1 \, \int_0^{\infty} d \omega_2 \,
{\rm exp} \left [- i (\omega_1 \tau_1 + \omega_2 \tau_2) \right ] \,
\Phi_{\rm G}(\omega_1, \omega_2, \mu)\,,
\label{def: phi_G}
\end{eqnarray}
\end{widetext}
where the two soft Wilson lines $S_{n}$ and $S_{\bar n}$ essential to
maintain  gauge invariance are given by
\begin{eqnarray}
S_{n}(x) &=& {\rm P \, exp} \left [ i \, g_s \int_{-\infty}^{0}  d t \, n \cdot A_s(x + t \, n) \right ] \,,
\nonumber \\
S_{\bar n}(x) &=& {\rm P \, exp} \left [ i \, g_s \int_{-\infty}^{0}  d t \, \bar n \cdot A_s(x + t \, \bar n) \right ]  \,.
\end{eqnarray}
It is apparent that the non-local HQET matrix element on the left-hand side of  (\ref{def: phi_G})
can be described by the familiar three-particle light-cone distribution amplitude when taking the limit $\tau_{1(2)} \to 0$.
We are therefore led to the following three important and model-independent normalization conditions  at tree level
\begin{eqnarray}
&& \int_0^{\infty} d \omega_1 \,  \Phi_{\rm G}(\omega_1, \omega_2, \mu)  =
\int_0^{\infty} d \omega_1 \,  \Phi_{4}(\omega_1, \omega_2, \mu) \,,
\nonumber \\
&& \int_0^{\infty} d \omega_2 \,  \Phi_{\rm G}(\omega_1, \omega_2, \mu)  =
\int_0^{\infty} d \omega_2 \,  \Phi_{5}(\omega_1, \omega_2, \mu) \,,
\nonumber \\
&& \int_0^{\infty} d \omega_1 \, \int_0^{\infty} d \omega_2 \,  \Phi_{\rm G}(\omega_1, \omega_2, \mu)  =
\frac{\lambda_E^2 + \lambda_H^2}{3}  \,,
\label{normalization conditions}
\end{eqnarray}
where the explicit definitions of the twist-four and twist-five light-cone distribution amplitudes
$\Phi_{4}$ and $\Phi_{5}$ can be found in \cite{Braun:2017liq} and the hadronic quantities ${\lambda}_E^2$
and ${\lambda}_H^2$ can be defined by the effective matrix elements of
the local chromoelectric and chromomagnetic operators \cite{Grozin:1996pq}.
Furthermore, the asymptotic behaviour of $\Phi_{\rm G}$ at small quark and gluon momenta can be
predicted with the dispersion technique as widely adopted in the explorations  of
the two-particle and three-particle  $B$-meson light-cone distribution amplitudes \cite{Grozin:1996pq,Braun:2003wx,Khodjamirian:2006st,Lu:2018cfc,Khodjamirian:2020hob}.
Starting with the HQET correlation function
\begin{eqnarray}
&& \Pi_{\rm G}
=  i \, \int d^4 x \, {\rm exp} \left (- i \, \omega \, v \cdot x \right ) \,
\langle 0 | {\rm T} \big  \{ \big [ (\bar q_{s} S_{n}) (\tau_1 n)
\nonumber \\
&& \,\, (S_{n}^{\dagger} \, S_{\bar n})(0) \,
(S_{\bar n}^{\dagger} \, g_s \, G_{\mu \nu} \, S_{\bar n} )(\tau_2 \bar n) \,\,
\bar n^{\nu} \not \! n \gamma_{\perp}^{\mu} \gamma_5 \,
(S_{\bar n}^{\dagger} h_v) (0) \big ],
\nonumber \\
&& \,\, \left [  \bar h_v(x) \, \, g_s \, G_{\rho \lambda}(x) \,
\sigma^{\rho \lambda} \, \gamma_5 \, q_s(x) \right ] \big  \} | 0 \rangle \,,
\end{eqnarray}
we can on the one hand compute this quantity in the kinematic region
$|\omega| \gg \Lambda$ with the operator-product-expansion (OPE) method
and on the other hand derive the corresponding hadronic representation of $\Pi_{\rm G}$
by taking advantage of analyticity with respect to the effective variable $\omega$.
Matching the above two dispersion representations with the aid of  the parton-hadron duality ansatz
enables us to extract the desired asymptotic behaviour
$\Phi_{\rm G}(\omega_1, \omega_2, \mu) \sim \omega_1 \, \omega_2^2$
at $\omega_1, \, \omega_2 \to 0$ immediately,
which further indicates that the convolution integrals in the factorized expression
(\ref{LO factorization formula}) converge.
We restrict ourselves to the leading-order accuracy in this letter, however,
it will be of interest to investigate the  non-trivial impact of the renormalization-group evolution
on the generalized  distribution amplitude $\Phi_{\rm G}$ in the future.

%
\section{Numerical implications}
%

We now turn to address the phenomenological implications of the soft-gluon radiative correction
to the penguin contractions of the effective four-quark operators on the double radiative
$\bar B_q \to \gamma \gamma$ decay amplitudes. To achieve this goal, we first need to construct
the acceptable non-perturbative model for the subleading distribution amplitude $\Phi_{\rm G}$
fulfilling the third relation in  (\ref{normalization conditions})
as well as the obtained asymptotic behaviour
\begin{eqnarray}
&& \Phi_{\rm G}(\omega_1, \omega_2, \mu_0) = \frac{\lambda_E^2 + \lambda_H^2}{6} \,
{\omega_1 \omega_2^2  \over \omega_0^5} \,
{\rm exp} \left (- {\omega_1 + \omega_2 \over \omega_0} \right )
\nonumber \\
&&  \hspace{1.5 cm} \frac{\Gamma(\beta+2)} {\Gamma(\alpha+2)} \,\,
U \left (\beta-\alpha, 4-\alpha, {\omega_1 + \omega_2 \over \omega_0} \right ) \,,
\end{eqnarray}
at the reference scale $\mu_0=1.0 \, {\rm GeV}$, motivated from the suggested three-parameter ansatz
for the twist-two distribution amplitude $\phi_B^{+}(\omega, \mu_0)$ \cite{Beneke:2018wjp}.
The remaining shape parameters $\omega_0$, $\alpha$ and $\beta$ can be further determined by enforcing  the
first and second normalization relations in (\ref{normalization conditions}) and employing the concrete model of
the twist-four and twist-five light-cone distribution amplitudes $\Phi_{4, \, 5}(\omega_1, \omega_2, \mu_0)$
as implemented in \cite{Gao:2021sav}.
The conventional HQET distribution amplitudes on the light-cone appearing in the established factorization formulae
of the helicity form factors $F_{i, L (R)}^{ (p)}$  from \cite{Shen:2020hfq} will be also in demand
in the subsequent numerical investigations, and we will apply the same phenomenological model
as presented in this reference, with the exceptions of  updated intervals for the inverse moments
$\lambda_{B_d} = (275 \pm 75)  \, {\rm MeV}$ and $\lambda_{B_s} = (325 \pm 75)  \, {\rm MeV}$ \cite{Beneke:2017vpq}
(see \cite{Khodjamirian:2020hob} for a recent determination of $\lambda_{B_s} / \lambda_{B_d}$ with the method of QCD sum rules).
The allowed intervals of additional theory input parameters entering our numerical studies are identical to the ones
collected in \cite{Shen:2020hfq}.


\begin{figure}[tp]
\begin{center}
\includegraphics[width=0.75 \columnwidth]{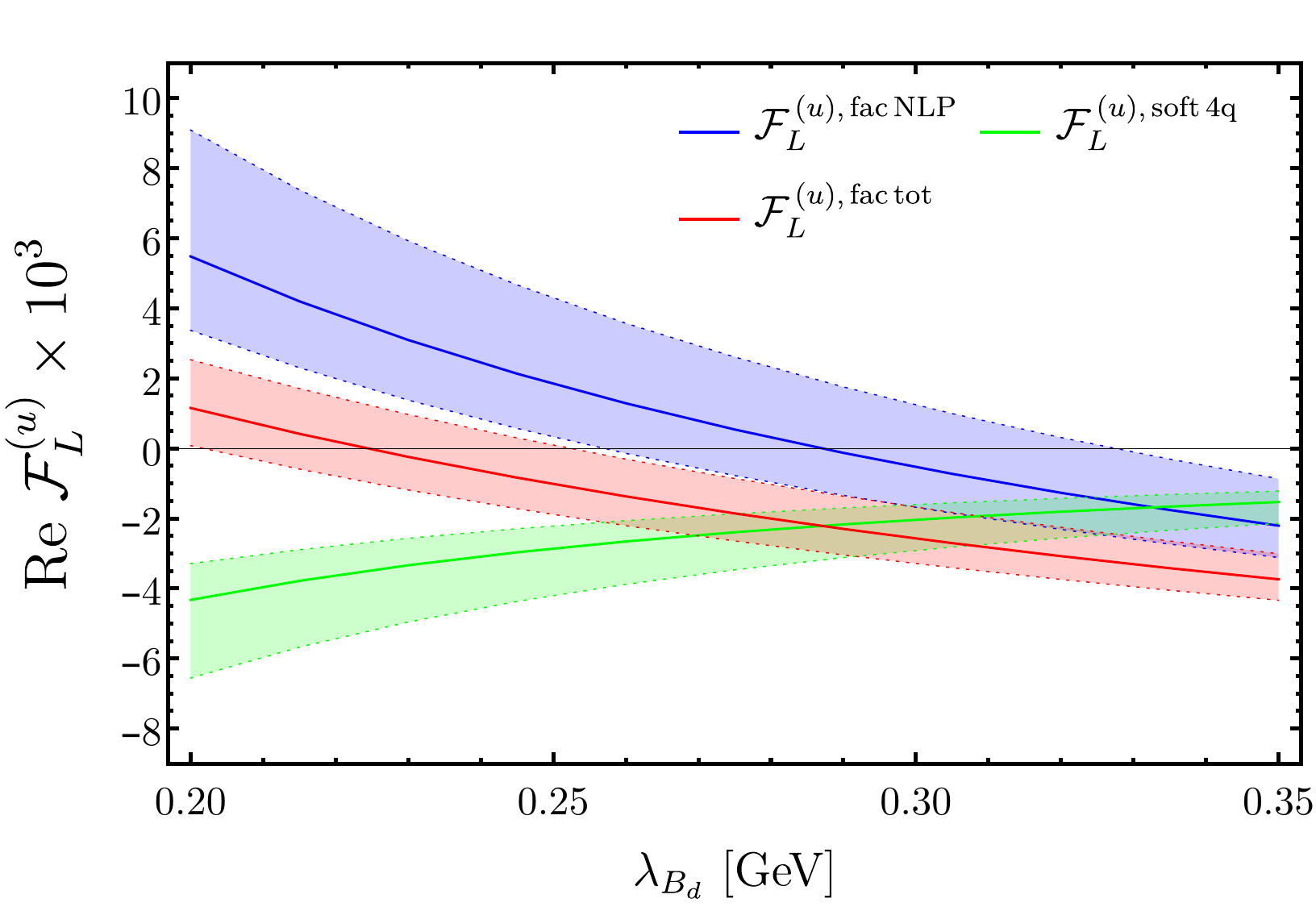}
\includegraphics[width=0.75 \columnwidth]{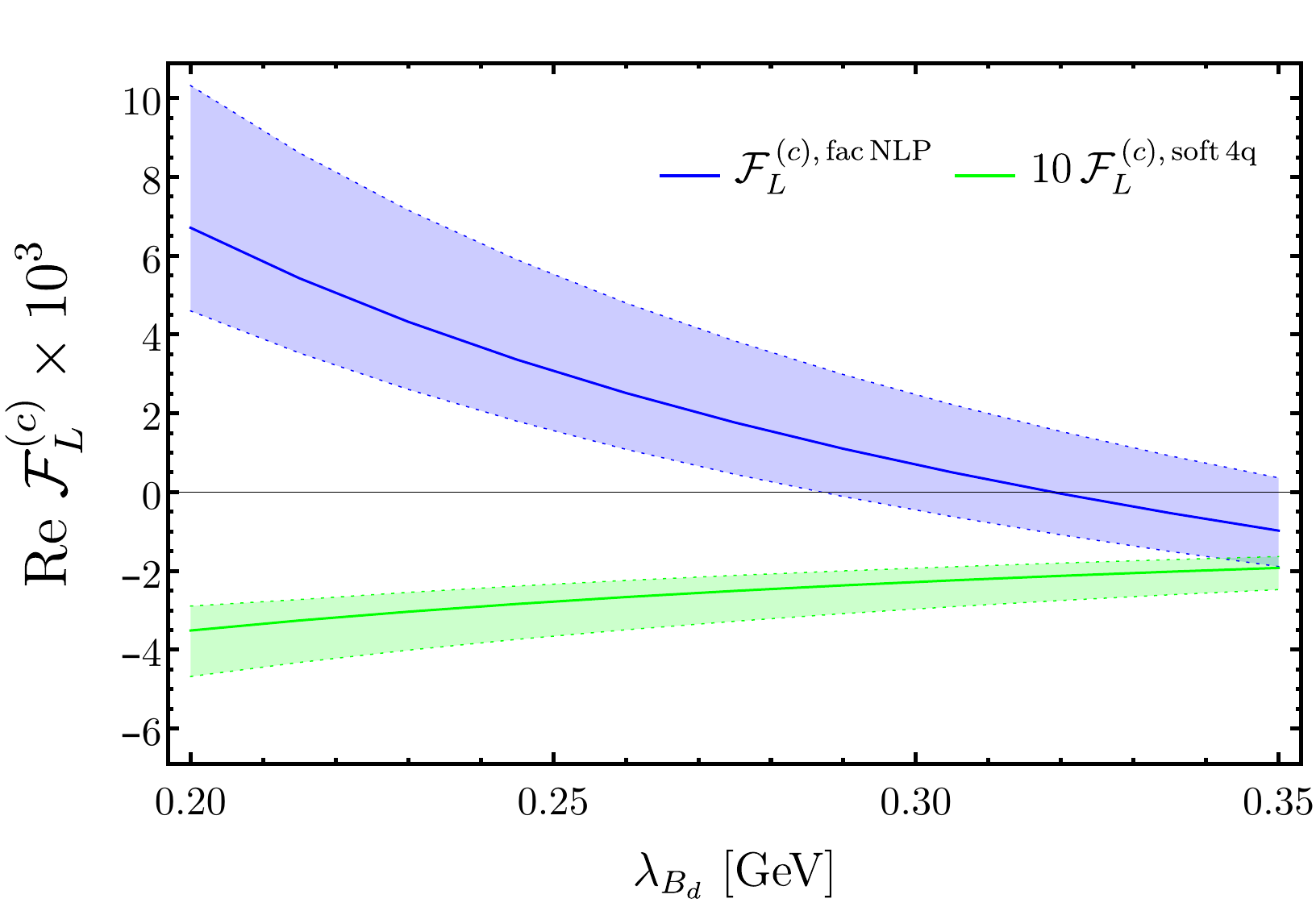}
\caption{Theory predictions for the soft-gluon radiative corrections to
the left-handed helicity form factors of $\bar B_q \to \gamma \gamma$
with the perturbative uncertainties from varying the factorization scale in the interval
$\mu \in [1.0, \, 2.0] \, {\rm GeV}$ (green bands),
where we further display the numerical results for the combined factorizable power corrections
as previously derived in \cite{Shen:2020hfq} (blue bands) for a comparison.
The shorthand notations for the weighted helicity form factors ${\cal F}_{L}^{(p), X} = \sum \limits_{i} \,  C_i \, F_{i, L }^{(p), X}$
have been introduced here for convenience. }
\label{fig: numerical results of helicity form factors}
\end{center}
\end{figure}

We  present the theory predictions for the long-distance penguin contribution
to the double radiative $B$-meson decay form factors including the obtained uncertainties
from varying the factorization scale $\mu$ in Figure \ref{fig: numerical results of helicity form factors},
where the numerical predictions of the previously computed factorizable power corrections
at tree level \cite{Shen:2020hfq} are also presented for the illustration purpose.
Interestingly, the newly computed soft gluon radiation from the up-quark penguin contraction appears to generate
the substantial cancellation of the combined factorizable  power corrections derived in \cite{Shen:2020hfq}.
On the contrary, the long-distance charm-quark penguin  mechanism will only lead to the rather minor impact on
the left-handed helicity form factor $\sum \limits_{i} \, C_i \, F_{i, L }^{ (c)}$ numerically
as indicated by Figure \ref{fig: numerical results of helicity form factors}.
This intriguing observation can be attributed to the peculiar analytical behaviour of
the perturbative penguin function $F(z_{p})$ entering the soft-collinear factorization formula
(\ref{LO factorization formula}) such that the yielding result of $\left | F \left (- m_c^2 /(m_B \, \omega_2) \right)-1 \right |$
in the bulk of the integral domain of $\omega_2$ is approximately one order of magnitude lower than
the corresponding up-quark penguin contribution $\left | F (0)-1 \right | = 1$.
Along the same vein, one can readily observe  that the  soft-gluon radiative corrections to  the factorizable quark loops
cannot generate numerically important contributions to the exclusive $\bar B_s \to \gamma \gamma$ helicity amplitudes
due to the CKM suppression of the up-quark penguin contraction  and the dynamical suppression of the charm-quark penguin
contribution as discussed above.
Bearing in mind the utmost importance of understanding the  charming penguin contribution in unveiling the genuine
NP effects embedded in the semileptonic $B \to K^{(\ast)} \ell \bar \ell$  decays
(see for instance \cite{Cerri:2018ypt,Ciuchini:2021smi,Altmannshofer:2022hfs}),
the achieved  robust  control  of such long-distance penguin contribution
in the double radiative $B$-meson decays with the diagrammatic factorization technique
evidently makes these FCNC  decay processes most suitable for probing the nonstandard
four-fermion $b \to q \, f \bar f$ interaction at the high-luminosity  Belle II experiment \cite{Belle-II:2018jsg}
and for performing the dedicated  parton tomography of the composite heavy-quark hadron system in the QCD framework.

We are now ready to explore the numerical impacts of the long-distance penguin contribution
on the CP-averaged branching fractions, the two polarization fractions and the CP-violating observables
for $\bar B_q \to \gamma \gamma$ (see \cite{Shen:2020hfq} for their explicit definitions).
It turns out that such subleading power corrections can enhance the theory predictions
for  the mixing induced CP asymmetries ${\cal A}_{\rm CP}^{\rm mix, \, \|}$ and ${\cal A}_{\rm CP}^{\rm mix, \, \perp}$
by approximately an amount of ${\cal O} (30 \, \%)$ with the default inputs,
while yielding  insignificant effects in the remaining observables numerically.
Moreover, our numerical prediction for the  ratio of the two branching fractions
${\cal BR}(B_s \to \gamma \gamma) : {\cal BR}(B_d \to \gamma \gamma)$ allows for extracting
the high-profile hadronic quantity $\lambda_{B_s} : \lambda_{B_d}$ with the improved systematic uncertainty
at the level of $ (5 - 10) \%$, when confronting with the anticipated precision measurements.

%
\section{Conclusions}
%

In conclusion, we have presented the first computation of the long-distance penguin contribution
to the double radiative $B$-meson decay amplitudes by applying the perturbative factorization approach.
Adopting the power counting scheme $m_b \gg m_c \sim {\cal O} (\sqrt{\Lambda \, m_b}) \gg \Lambda$,
we demonstrated further that the novel subleading $B$-meson distribution amplitude $\Phi_{\rm G}$
defined by the three-body HQET operator with partonic fields localized on two different light-ray directions
(instead of the conventional light-cone distribution amplitude)
emerged naturally in the resulting factorization formula (\ref{LO factorization formula}).
Phenomenologically the soft-gluon radiative off the factorizable up-quark loop appeared to
bring about  more pronounced effect in comparison with the corresponding charm-quark penguin mechanism
thanks to the peculiar analytical behaviour of the short-distance matching function (\ref{penguin function}).
In addition, the observed destructive interference  between the long-distance penguin contribution
and the available factorizable power correction from \cite{Shen:2020hfq} enabled us to determine
the pivotal ratio of the two inverse moments $\lambda_{B_{d, s}}$ with the reduced theory uncertainty.
Our analysis will be evidently beneficial for exploring the intricate charming penguin dynamics
encoded in a large variety of the radiative and electroweak penguin decay processes
including $B \to K^{\ast} \gamma$ and $B \to K^{(\ast)} \ell \bar \ell$,
which are generally recognized as the flagship probes of physics beyond the SM at the LHC.

%
\begin{acknowledgments}
\section*{Acknowledgements}

The research of Q.Q. is supported by the National Natural
Science Foundation of China with Grant No. 12005068.
C.W. is supported in part by the National Natural Science Foundation of China
with Grant No. 12105112 and  the Natural Science Foundation of
Jiangsu Education Committee with Grant No. 21KJB140027.
The research of Y.L.S. is supported by the  National Natural Science Foundation of China  with
Grant No. 12175218 and the Natural Science Foundation of Shandong with Grant No.  ZR2020MA093.
Y.M.W. acknowledges support from the  National Natural Science Foundation of China  with
Grant No. 11735010 and 12075125, and the Natural Science Foundation of Tianjin
with Grant No. 19JCJQJC61100.

\end{acknowledgments}

\bibliographystyle{apsrev4-1}

\bibliography{References}

\begin{thebibliography}{61}%
\makeatletter
\providecommand \@ifxundefined [1]{%
 \@ifx{#1\undefined}
}%
\providecommand \@ifnum [1]{%
 \ifnum #1\expandafter \@firstoftwo
 \else \expandafter \@secondoftwo
 \fi
}%
\providecommand \@ifx [1]{%
 \ifx #1\expandafter \@firstoftwo
 \else \expandafter \@secondoftwo
 \fi
}%
\providecommand \natexlab [1]{#1}%
\providecommand \enquote  [1]{``#1''}%
\providecommand \bibnamefont  [1]{#1}%
\providecommand \bibfnamefont [1]{#1}%
\providecommand \citenamefont [1]{#1}%
\providecommand \href@noop [0]{\@secondoftwo}%
\providecommand \href [0]{\begingroup \@sanitize@url \@href}%
\providecommand \@href[1]{\@@startlink{#1}\@@href}%
\providecommand \@@href[1]{\endgroup#1\@@endlink}%
\providecommand \@sanitize@url [0]{\catcode `\\12\catcode `\$12\catcode
  `\&12\catcode `\#12\catcode `\^12\catcode `\_12\catcode `\%12\relax}%
\providecommand \@@startlink[1]{}%
\providecommand \@@endlink[0]{}%
\providecommand \url  [0]{\begingroup\@sanitize@url \@url }%
\providecommand \@url [1]{\endgroup\@href {#1}{\urlprefix }}%
\providecommand \urlprefix  [0]{URL }%
\providecommand \Eprint [0]{\href }%
\providecommand \doibase [0]{http://dx.doi.org/}%
\providecommand \selectlanguage [0]{\@gobble}%
\providecommand \bibinfo  [0]{\@secondoftwo}%
\providecommand \bibfield  [0]{\@secondoftwo}%
\providecommand \translation [1]{[#1]}%
\providecommand \BibitemOpen [0]{}%
\providecommand \bibitemStop [0]{}%
\providecommand \bibitemNoStop [0]{.\EOS\space}%
\providecommand \EOS [0]{\spacefactor3000\relax}%
\providecommand \BibitemShut  [1]{\csname bibitem#1\endcsname}%
\let\auto@bib@innerbib\@empty
\bibitem [{\citenamefont {Bosch}\ and\ \citenamefont
  {Buchalla}(2002)}]{Bosch:2002bv}%
  \BibitemOpen
  \bibfield  {author} {\bibinfo {author} {\bibfnamefont {S.~W.}\ \bibnamefont
  {Bosch}}\ and\ \bibinfo {author} {\bibfnamefont {G.}~\bibnamefont
  {Buchalla}},\ }\href {\doibase 10.1088/1126-6708/2002/08/054} {\bibfield
  {journal} {\bibinfo  {journal} {JHEP}\ }\textbf {\bibinfo {volume} {08}},\
  \bibinfo {pages} {054} (\bibinfo {year} {2002})},\ \Eprint
  {http://arxiv.org/abs/hep-ph/0208202} {arXiv:hep-ph/0208202} \BibitemShut
  {NoStop}%
\bibitem [{\citenamefont {Descotes-Genon}\ and\ \citenamefont
  {Sachrajda}(2003)}]{Descotes-Genon:2002lal}%
  \BibitemOpen
  \bibfield  {author} {\bibinfo {author} {\bibfnamefont {S.}~\bibnamefont
  {Descotes-Genon}}\ and\ \bibinfo {author} {\bibfnamefont {C.~T.}\
  \bibnamefont {Sachrajda}},\ }\href {\doibase 10.1016/S0370-2693(03)00173-4}
  {\bibfield  {journal} {\bibinfo  {journal} {Phys. Lett. B}\ }\textbf
  {\bibinfo {volume} {557}},\ \bibinfo {pages} {213} (\bibinfo {year}
  {2003})},\ \Eprint {http://arxiv.org/abs/hep-ph/0212162}
  {arXiv:hep-ph/0212162} \BibitemShut {NoStop}%
\bibitem [{\citenamefont {Shen}\ \emph {et~al.}(2020)\citenamefont {Shen},
  \citenamefont {Wang},\ and\ \citenamefont {Wei}}]{Shen:2020hfq}%
  \BibitemOpen
  \bibfield  {author} {\bibinfo {author} {\bibfnamefont {Y.-L.}\ \bibnamefont
  {Shen}}, \bibinfo {author} {\bibfnamefont {Y.-M.}\ \bibnamefont {Wang}}, \
  and\ \bibinfo {author} {\bibfnamefont {Y.-B.}\ \bibnamefont {Wei}},\ }\href
  {\doibase 10.1007/JHEP12(2020)169} {\bibfield  {journal} {\bibinfo  {journal}
  {JHEP}\ }\textbf {\bibinfo {volume} {12}},\ \bibinfo {pages} {169} (\bibinfo
  {year} {2020})},\ \Eprint {http://arxiv.org/abs/2009.02723} {arXiv:2009.02723
  [hep-ph]} \BibitemShut {NoStop}%
\bibitem [{\citenamefont {Ali}\ \emph {et~al.}(1991)\citenamefont {Ali},
  \citenamefont {Mannel},\ and\ \citenamefont {Morozumi}}]{Ali:1991is}%
  \BibitemOpen
  \bibfield  {author} {\bibinfo {author} {\bibfnamefont {A.}~\bibnamefont
  {Ali}}, \bibinfo {author} {\bibfnamefont {T.}~\bibnamefont {Mannel}}, \ and\
  \bibinfo {author} {\bibfnamefont {T.}~\bibnamefont {Morozumi}},\ }\href
  {\doibase 10.1016/0370-2693(91)90306-B} {\bibfield  {journal} {\bibinfo
  {journal} {Phys. Lett. B}\ }\textbf {\bibinfo {volume} {273}},\ \bibinfo
  {pages} {505} (\bibinfo {year} {1991})}\BibitemShut {NoStop}%
\bibitem [{\citenamefont {Khodjamirian}\ \emph {et~al.}(2010)\citenamefont
  {Khodjamirian}, \citenamefont {Mannel}, \citenamefont {Pivovarov},\ and\
  \citenamefont {Wang}}]{Khodjamirian:2010vf}%
  \BibitemOpen
  \bibfield  {author} {\bibinfo {author} {\bibfnamefont {A.}~\bibnamefont
  {Khodjamirian}}, \bibinfo {author} {\bibfnamefont {T.}~\bibnamefont
  {Mannel}}, \bibinfo {author} {\bibfnamefont {A.~A.}\ \bibnamefont
  {Pivovarov}}, \ and\ \bibinfo {author} {\bibfnamefont {Y.~M.}\ \bibnamefont
  {Wang}},\ }\href {\doibase 10.1007/JHEP09(2010)089} {\bibfield  {journal}
  {\bibinfo  {journal} {JHEP}\ }\textbf {\bibinfo {volume} {09}},\ \bibinfo
  {pages} {089} (\bibinfo {year} {2010})},\ \Eprint
  {http://arxiv.org/abs/1006.4945} {arXiv:1006.4945 [hep-ph]} \BibitemShut
  {NoStop}%
\bibitem [{\citenamefont {Khodjamirian}\ \emph {et~al.}(2013)\citenamefont
  {Khodjamirian}, \citenamefont {Mannel},\ and\ \citenamefont
  {Wang}}]{Khodjamirian:2012rm}%
  \BibitemOpen
  \bibfield  {author} {\bibinfo {author} {\bibfnamefont {A.}~\bibnamefont
  {Khodjamirian}}, \bibinfo {author} {\bibfnamefont {T.}~\bibnamefont
  {Mannel}}, \ and\ \bibinfo {author} {\bibfnamefont {Y.~M.}\ \bibnamefont
  {Wang}},\ }\href {\doibase 10.1007/JHEP02(2013)010} {\bibfield  {journal}
  {\bibinfo  {journal} {JHEP}\ }\textbf {\bibinfo {volume} {02}},\ \bibinfo
  {pages} {010} (\bibinfo {year} {2013})},\ \Eprint
  {http://arxiv.org/abs/1211.0234} {arXiv:1211.0234 [hep-ph]} \BibitemShut
  {NoStop}%
\bibitem [{\citenamefont {Hambrock}\ \emph {et~al.}(2015)\citenamefont
  {Hambrock}, \citenamefont {Khodjamirian},\ and\ \citenamefont
  {Rusov}}]{Hambrock:2015wka}%
  \BibitemOpen
  \bibfield  {author} {\bibinfo {author} {\bibfnamefont {C.}~\bibnamefont
  {Hambrock}}, \bibinfo {author} {\bibfnamefont {A.}~\bibnamefont
  {Khodjamirian}}, \ and\ \bibinfo {author} {\bibfnamefont {A.}~\bibnamefont
  {Rusov}},\ }\href {\doibase 10.1103/PhysRevD.92.074020} {\bibfield  {journal}
  {\bibinfo  {journal} {Phys. Rev. D}\ }\textbf {\bibinfo {volume} {92}},\
  \bibinfo {pages} {074020} (\bibinfo {year} {2015})},\ \Eprint
  {http://arxiv.org/abs/1506.07760} {arXiv:1506.07760 [hep-ph]} \BibitemShut
  {NoStop}%
\bibitem [{\citenamefont {Bobeth}\ \emph {et~al.}(2018)\citenamefont {Bobeth},
  \citenamefont {Chrzaszcz}, \citenamefont {van Dyk},\ and\ \citenamefont
  {Virto}}]{Bobeth:2017vxj}%
  \BibitemOpen
  \bibfield  {author} {\bibinfo {author} {\bibfnamefont {C.}~\bibnamefont
  {Bobeth}}, \bibinfo {author} {\bibfnamefont {M.}~\bibnamefont {Chrzaszcz}},
  \bibinfo {author} {\bibfnamefont {D.}~\bibnamefont {van Dyk}}, \ and\
  \bibinfo {author} {\bibfnamefont {J.}~\bibnamefont {Virto}},\ }\href
  {\doibase 10.1140/epjc/s10052-018-5918-6} {\bibfield  {journal} {\bibinfo
  {journal} {Eur. Phys. J. C}\ }\textbf {\bibinfo {volume} {78}},\ \bibinfo
  {pages} {451} (\bibinfo {year} {2018})},\ \Eprint
  {http://arxiv.org/abs/1707.07305} {arXiv:1707.07305 [hep-ph]} \BibitemShut
  {NoStop}%
\bibitem [{\citenamefont {Kozachuk}\ and\ \citenamefont
  {Melikhov}(2018)}]{Kozachuk:2018yxf}%
  \BibitemOpen
  \bibfield  {author} {\bibinfo {author} {\bibfnamefont {A.}~\bibnamefont
  {Kozachuk}}\ and\ \bibinfo {author} {\bibfnamefont {D.}~\bibnamefont
  {Melikhov}},\ }\href {\doibase 10.1016/j.physletb.2018.10.026} {\bibfield
  {journal} {\bibinfo  {journal} {Phys. Lett. B}\ }\textbf {\bibinfo {volume}
  {786}},\ \bibinfo {pages} {378} (\bibinfo {year} {2018})},\ \Eprint
  {http://arxiv.org/abs/1805.05720} {arXiv:1805.05720 [hep-ph]} \BibitemShut
  {NoStop}%
\bibitem [{\citenamefont {Melikhov}(2019)}]{Melikhov:2019esw}%
  \BibitemOpen
  \bibfield  {author} {\bibinfo {author} {\bibfnamefont {D.}~\bibnamefont
  {Melikhov}},\ }\href {\doibase 10.1051/epjconf/201922201007} {\bibfield
  {journal} {\bibinfo  {journal} {EPJ Web Conf.}\ }\textbf {\bibinfo {volume}
  {222}},\ \bibinfo {pages} {01007} (\bibinfo {year} {2019})},\ \Eprint
  {http://arxiv.org/abs/1911.03899} {arXiv:1911.03899 [hep-ph]} \BibitemShut
  {NoStop}%
\bibitem [{\citenamefont {Gubernari}\ \emph {et~al.}(2021)\citenamefont
  {Gubernari}, \citenamefont {van Dyk},\ and\ \citenamefont
  {Virto}}]{Gubernari:2020eft}%
  \BibitemOpen
  \bibfield  {author} {\bibinfo {author} {\bibfnamefont {N.}~\bibnamefont
  {Gubernari}}, \bibinfo {author} {\bibfnamefont {D.}~\bibnamefont {van Dyk}},
  \ and\ \bibinfo {author} {\bibfnamefont {J.}~\bibnamefont {Virto}},\ }\href
  {\doibase 10.1007/JHEP02(2021)088} {\bibfield  {journal} {\bibinfo  {journal}
  {JHEP}\ }\textbf {\bibinfo {volume} {02}},\ \bibinfo {pages} {088} (\bibinfo
  {year} {2021})},\ \Eprint {http://arxiv.org/abs/2011.09813} {arXiv:2011.09813
  [hep-ph]} \BibitemShut {NoStop}%
\bibitem [{\citenamefont {Grinstein}\ and\ \citenamefont
  {Pirjol}(2004)}]{Grinstein:2004vb}%
  \BibitemOpen
  \bibfield  {author} {\bibinfo {author} {\bibfnamefont {B.}~\bibnamefont
  {Grinstein}}\ and\ \bibinfo {author} {\bibfnamefont {D.}~\bibnamefont
  {Pirjol}},\ }\href {\doibase 10.1103/PhysRevD.70.114005} {\bibfield
  {journal} {\bibinfo  {journal} {Phys. Rev. D}\ }\textbf {\bibinfo {volume}
  {70}},\ \bibinfo {pages} {114005} (\bibinfo {year} {2004})},\ \Eprint
  {http://arxiv.org/abs/hep-ph/0404250} {arXiv:hep-ph/0404250} \BibitemShut
  {NoStop}%
\bibitem [{\citenamefont {Beylich}\ \emph {et~al.}(2011)\citenamefont
  {Beylich}, \citenamefont {Buchalla},\ and\ \citenamefont
  {Feldmann}}]{Beylich:2011aq}%
  \BibitemOpen
  \bibfield  {author} {\bibinfo {author} {\bibfnamefont {M.}~\bibnamefont
  {Beylich}}, \bibinfo {author} {\bibfnamefont {G.}~\bibnamefont {Buchalla}}, \
  and\ \bibinfo {author} {\bibfnamefont {T.}~\bibnamefont {Feldmann}},\ }\href
  {\doibase 10.1140/epjc/s10052-011-1635-0} {\bibfield  {journal} {\bibinfo
  {journal} {Eur. Phys. J. C}\ }\textbf {\bibinfo {volume} {71}},\ \bibinfo
  {pages} {1635} (\bibinfo {year} {2011})},\ \Eprint
  {http://arxiv.org/abs/1101.5118} {arXiv:1101.5118 [hep-ph]} \BibitemShut
  {NoStop}%
\bibitem [{\citenamefont {Lyon}\ and\ \citenamefont
  {Zwicky}(2014)}]{Lyon:2014hpa}%
  \BibitemOpen
  \bibfield  {author} {\bibinfo {author} {\bibfnamefont {J.}~\bibnamefont
  {Lyon}}\ and\ \bibinfo {author} {\bibfnamefont {R.}~\bibnamefont {Zwicky}},\
  }\href@noop {} {\  (\bibinfo {year} {2014})},\ \Eprint
  {http://arxiv.org/abs/1406.0566} {arXiv:1406.0566 [hep-ph]} \BibitemShut
  {NoStop}%
\bibitem [{\citenamefont {J\"ager}\ and\ \citenamefont
  {Martin~Camalich}(2013)}]{Jager:2012uw}%
  \BibitemOpen
  \bibfield  {author} {\bibinfo {author} {\bibfnamefont {S.}~\bibnamefont
  {J\"ager}}\ and\ \bibinfo {author} {\bibfnamefont {J.}~\bibnamefont
  {Martin~Camalich}},\ }\href {\doibase 10.1007/JHEP05(2013)043} {\bibfield
  {journal} {\bibinfo  {journal} {JHEP}\ }\textbf {\bibinfo {volume} {05}},\
  \bibinfo {pages} {043} (\bibinfo {year} {2013})},\ \Eprint
  {http://arxiv.org/abs/1212.2263} {arXiv:1212.2263 [hep-ph]} \BibitemShut
  {NoStop}%
\bibitem [{\citenamefont {Descotes-Genon}\ \emph {et~al.}(2013)\citenamefont
  {Descotes-Genon}, \citenamefont {Matias},\ and\ \citenamefont
  {Virto}}]{Descotes-Genon:2013wba}%
  \BibitemOpen
  \bibfield  {author} {\bibinfo {author} {\bibfnamefont {S.}~\bibnamefont
  {Descotes-Genon}}, \bibinfo {author} {\bibfnamefont {J.}~\bibnamefont
  {Matias}}, \ and\ \bibinfo {author} {\bibfnamefont {J.}~\bibnamefont
  {Virto}},\ }\href {\doibase 10.1103/PhysRevD.88.074002} {\bibfield  {journal}
  {\bibinfo  {journal} {Phys. Rev. D}\ }\textbf {\bibinfo {volume} {88}},\
  \bibinfo {pages} {074002} (\bibinfo {year} {2013})},\ \Eprint
  {http://arxiv.org/abs/1307.5683} {arXiv:1307.5683 [hep-ph]} \BibitemShut
  {NoStop}%
\bibitem [{\citenamefont {Descotes-Genon}\ \emph {et~al.}(2016)\citenamefont
  {Descotes-Genon}, \citenamefont {Hofer}, \citenamefont {Matias},\ and\
  \citenamefont {Virto}}]{Descotes-Genon:2015uva}%
  \BibitemOpen
  \bibfield  {author} {\bibinfo {author} {\bibfnamefont {S.}~\bibnamefont
  {Descotes-Genon}}, \bibinfo {author} {\bibfnamefont {L.}~\bibnamefont
  {Hofer}}, \bibinfo {author} {\bibfnamefont {J.}~\bibnamefont {Matias}}, \
  and\ \bibinfo {author} {\bibfnamefont {J.}~\bibnamefont {Virto}},\ }\href
  {\doibase 10.1007/JHEP06(2016)092} {\bibfield  {journal} {\bibinfo  {journal}
  {JHEP}\ }\textbf {\bibinfo {volume} {06}},\ \bibinfo {pages} {092} (\bibinfo
  {year} {2016})},\ \Eprint {http://arxiv.org/abs/1510.04239} {arXiv:1510.04239
  [hep-ph]} \BibitemShut {NoStop}%
\bibitem [{\citenamefont {Ciuchini}\ \emph {et~al.}(2016)\citenamefont
  {Ciuchini}, \citenamefont {Fedele}, \citenamefont {Franco}, \citenamefont
  {Mishima}, \citenamefont {Paul}, \citenamefont {Silvestrini},\ and\
  \citenamefont {Valli}}]{Ciuchini:2015qxb}%
  \BibitemOpen
  \bibfield  {author} {\bibinfo {author} {\bibfnamefont {M.}~\bibnamefont
  {Ciuchini}}, \bibinfo {author} {\bibfnamefont {M.}~\bibnamefont {Fedele}},
  \bibinfo {author} {\bibfnamefont {E.}~\bibnamefont {Franco}}, \bibinfo
  {author} {\bibfnamefont {S.}~\bibnamefont {Mishima}}, \bibinfo {author}
  {\bibfnamefont {A.}~\bibnamefont {Paul}}, \bibinfo {author} {\bibfnamefont
  {L.}~\bibnamefont {Silvestrini}}, \ and\ \bibinfo {author} {\bibfnamefont
  {M.}~\bibnamefont {Valli}},\ }\href {\doibase 10.1007/JHEP06(2016)116}
  {\bibfield  {journal} {\bibinfo  {journal} {JHEP}\ }\textbf {\bibinfo
  {volume} {06}},\ \bibinfo {pages} {116} (\bibinfo {year} {2016})},\ \Eprint
  {http://arxiv.org/abs/1512.07157} {arXiv:1512.07157 [hep-ph]} \BibitemShut
  {NoStop}%
\bibitem [{\citenamefont {Aebischer}\ \emph {et~al.}(2020)\citenamefont
  {Aebischer}, \citenamefont {Altmannshofer}, \citenamefont {Guadagnoli},
  \citenamefont {Reboud}, \citenamefont {Stangl},\ and\ \citenamefont
  {Straub}}]{Aebischer:2019mlg}%
  \BibitemOpen
  \bibfield  {author} {\bibinfo {author} {\bibfnamefont {J.}~\bibnamefont
  {Aebischer}}, \bibinfo {author} {\bibfnamefont {W.}~\bibnamefont
  {Altmannshofer}}, \bibinfo {author} {\bibfnamefont {D.}~\bibnamefont
  {Guadagnoli}}, \bibinfo {author} {\bibfnamefont {M.}~\bibnamefont {Reboud}},
  \bibinfo {author} {\bibfnamefont {P.}~\bibnamefont {Stangl}}, \ and\ \bibinfo
  {author} {\bibfnamefont {D.~M.}\ \bibnamefont {Straub}},\ }\href {\doibase
  10.1140/epjc/s10052-020-7817-x} {\bibfield  {journal} {\bibinfo  {journal}
  {Eur. Phys. J. C}\ }\textbf {\bibinfo {volume} {80}},\ \bibinfo {pages} {252}
  (\bibinfo {year} {2020})},\ \Eprint {http://arxiv.org/abs/1903.10434}
  {arXiv:1903.10434 [hep-ph]} \BibitemShut {NoStop}%
\bibitem [{\citenamefont {Ciuchini}\ \emph
  {et~al.}(2021{\natexlab{a}})\citenamefont {Ciuchini}, \citenamefont {Fedele},
  \citenamefont {Franco}, \citenamefont {Paul}, \citenamefont {Silvestrini},\
  and\ \citenamefont {Valli}}]{Ciuchini:2020gvn}%
  \BibitemOpen
  \bibfield  {author} {\bibinfo {author} {\bibfnamefont {M.}~\bibnamefont
  {Ciuchini}}, \bibinfo {author} {\bibfnamefont {M.}~\bibnamefont {Fedele}},
  \bibinfo {author} {\bibfnamefont {E.}~\bibnamefont {Franco}}, \bibinfo
  {author} {\bibfnamefont {A.}~\bibnamefont {Paul}}, \bibinfo {author}
  {\bibfnamefont {L.}~\bibnamefont {Silvestrini}}, \ and\ \bibinfo {author}
  {\bibfnamefont {M.}~\bibnamefont {Valli}},\ }\href {\doibase
  10.1103/PhysRevD.103.015030} {\bibfield  {journal} {\bibinfo  {journal}
  {Phys. Rev. D}\ }\textbf {\bibinfo {volume} {103}},\ \bibinfo {pages}
  {015030} (\bibinfo {year} {2021}{\natexlab{a}})},\ \Eprint
  {http://arxiv.org/abs/2011.01212} {arXiv:2011.01212 [hep-ph]} \BibitemShut
  {NoStop}%
\bibitem [{\citenamefont {Chen}\ and\ \citenamefont
  {Geng}(2001)}]{Chen:2001zc}%
  \BibitemOpen
  \bibfield  {author} {\bibinfo {author} {\bibfnamefont {C.-H.}\ \bibnamefont
  {Chen}}\ and\ \bibinfo {author} {\bibfnamefont {C.~Q.}\ \bibnamefont
  {Geng}},\ }\href {\doibase 10.1103/PhysRevD.64.074001} {\bibfield  {journal}
  {\bibinfo  {journal} {Phys. Rev. D}\ }\textbf {\bibinfo {volume} {64}},\
  \bibinfo {pages} {074001} (\bibinfo {year} {2001})},\ \Eprint
  {http://arxiv.org/abs/hep-ph/0106193} {arXiv:hep-ph/0106193} \BibitemShut
  {NoStop}%
\bibitem [{\citenamefont {He}\ \emph {et~al.}(2006)\citenamefont {He},
  \citenamefont {Li}, \citenamefont {Li},\ and\ \citenamefont
  {Wang}}]{He:2006ud}%
  \BibitemOpen
  \bibfield  {author} {\bibinfo {author} {\bibfnamefont {X.-G.}\ \bibnamefont
  {He}}, \bibinfo {author} {\bibfnamefont {T.}~\bibnamefont {Li}}, \bibinfo
  {author} {\bibfnamefont {X.-Q.}\ \bibnamefont {Li}}, \ and\ \bibinfo {author}
  {\bibfnamefont {Y.-M.}\ \bibnamefont {Wang}},\ }\href {\doibase
  10.1103/PhysRevD.74.034026} {\bibfield  {journal} {\bibinfo  {journal} {Phys.
  Rev. D}\ }\textbf {\bibinfo {volume} {74}},\ \bibinfo {pages} {034026}
  (\bibinfo {year} {2006})},\ \Eprint {http://arxiv.org/abs/hep-ph/0606025}
  {arXiv:hep-ph/0606025} \BibitemShut {NoStop}%
\bibitem [{\citenamefont {Wang}\ \emph
  {et~al.}(2009{\natexlab{a}})\citenamefont {Wang}, \citenamefont {Li},\ and\
  \citenamefont {Lu}}]{Wang:2008sm}%
  \BibitemOpen
  \bibfield  {author} {\bibinfo {author} {\bibfnamefont {Y.-M.}\ \bibnamefont
  {Wang}}, \bibinfo {author} {\bibfnamefont {Y.}~\bibnamefont {Li}}, \ and\
  \bibinfo {author} {\bibfnamefont {C.-D.}\ \bibnamefont {Lu}},\ }\href
  {\doibase 10.1140/epjc/s10052-008-0846-5} {\bibfield  {journal} {\bibinfo
  {journal} {Eur. Phys. J. C}\ }\textbf {\bibinfo {volume} {59}},\ \bibinfo
  {pages} {861} (\bibinfo {year} {2009}{\natexlab{a}})},\ \Eprint
  {http://arxiv.org/abs/0804.0648} {arXiv:0804.0648 [hep-ph]} \BibitemShut
  {NoStop}%
\bibitem [{\citenamefont {Ball}\ \emph {et~al.}(2008)\citenamefont {Ball},
  \citenamefont {Braun},\ and\ \citenamefont {Gardi}}]{Ball:2008fw}%
  \BibitemOpen
  \bibfield  {author} {\bibinfo {author} {\bibfnamefont {P.}~\bibnamefont
  {Ball}}, \bibinfo {author} {\bibfnamefont {V.~M.}\ \bibnamefont {Braun}}, \
  and\ \bibinfo {author} {\bibfnamefont {E.}~\bibnamefont {Gardi}},\ }\href
  {\doibase 10.1016/j.physletb.2008.06.004} {\bibfield  {journal} {\bibinfo
  {journal} {Phys. Lett. B}\ }\textbf {\bibinfo {volume} {665}},\ \bibinfo
  {pages} {197} (\bibinfo {year} {2008})},\ \Eprint
  {http://arxiv.org/abs/0804.2424} {arXiv:0804.2424 [hep-ph]} \BibitemShut
  {NoStop}%
\bibitem [{\citenamefont {Wang}\ \emph
  {et~al.}(2009{\natexlab{b}})\citenamefont {Wang}, \citenamefont {Shen},\ and\
  \citenamefont {Lu}}]{Wang:2009hra}%
  \BibitemOpen
  \bibfield  {author} {\bibinfo {author} {\bibfnamefont {Y.-M.}\ \bibnamefont
  {Wang}}, \bibinfo {author} {\bibfnamefont {Y.-L.}\ \bibnamefont {Shen}}, \
  and\ \bibinfo {author} {\bibfnamefont {C.-D.}\ \bibnamefont {Lu}},\ }\href
  {\doibase 10.1103/PhysRevD.80.074012} {\bibfield  {journal} {\bibinfo
  {journal} {Phys. Rev. D}\ }\textbf {\bibinfo {volume} {80}},\ \bibinfo
  {pages} {074012} (\bibinfo {year} {2009}{\natexlab{b}})},\ \Eprint
  {http://arxiv.org/abs/0907.4008} {arXiv:0907.4008 [hep-ph]} \BibitemShut
  {NoStop}%
\bibitem [{\citenamefont {Mannel}\ and\ \citenamefont
  {Wang}(2011)}]{Mannel:2011xg}%
  \BibitemOpen
  \bibfield  {author} {\bibinfo {author} {\bibfnamefont {T.}~\bibnamefont
  {Mannel}}\ and\ \bibinfo {author} {\bibfnamefont {Y.-M.}\ \bibnamefont
  {Wang}},\ }\href {\doibase 10.1007/JHEP12(2011)067} {\bibfield  {journal}
  {\bibinfo  {journal} {JHEP}\ }\textbf {\bibinfo {volume} {12}},\ \bibinfo
  {pages} {067} (\bibinfo {year} {2011})},\ \Eprint
  {http://arxiv.org/abs/1111.1849} {arXiv:1111.1849 [hep-ph]} \BibitemShut
  {NoStop}%
\bibitem [{\citenamefont {Feldmann}\ and\ \citenamefont
  {Yip}(2012)}]{Feldmann:2011xf}%
  \BibitemOpen
  \bibfield  {author} {\bibinfo {author} {\bibfnamefont {T.}~\bibnamefont
  {Feldmann}}\ and\ \bibinfo {author} {\bibfnamefont {M.~W.~Y.}\ \bibnamefont
  {Yip}},\ }\href {\doibase 10.1103/PhysRevD.85.014035} {\bibfield  {journal}
  {\bibinfo  {journal} {Phys. Rev. D}\ }\textbf {\bibinfo {volume} {85}},\
  \bibinfo {pages} {014035} (\bibinfo {year} {2012})},\ \bibinfo {note}
  {[Erratum: Phys.Rev.D 86, 079901 (2012)]},\ \Eprint
  {http://arxiv.org/abs/1111.1844} {arXiv:1111.1844 [hep-ph]} \BibitemShut
  {NoStop}%
\bibitem [{\citenamefont {Wang}(2012)}]{Wang:2011uv}%
  \BibitemOpen
  \bibfield  {author} {\bibinfo {author} {\bibfnamefont {W.}~\bibnamefont
  {Wang}},\ }\href {\doibase 10.1016/j.physletb.2012.01.036} {\bibfield
  {journal} {\bibinfo  {journal} {Phys. Lett. B}\ }\textbf {\bibinfo {volume}
  {708}},\ \bibinfo {pages} {119} (\bibinfo {year} {2012})},\ \Eprint
  {http://arxiv.org/abs/1112.0237} {arXiv:1112.0237 [hep-ph]} \BibitemShut
  {NoStop}%
\bibitem [{\citenamefont {Braun}\ \emph {et~al.}(2014)\citenamefont {Braun},
  \citenamefont {Derkachov},\ and\ \citenamefont {Manashov}}]{Braun:2014npa}%
  \BibitemOpen
  \bibfield  {author} {\bibinfo {author} {\bibfnamefont {V.~M.}\ \bibnamefont
  {Braun}}, \bibinfo {author} {\bibfnamefont {S.~E.}\ \bibnamefont
  {Derkachov}}, \ and\ \bibinfo {author} {\bibfnamefont {A.~N.}\ \bibnamefont
  {Manashov}},\ }\href {\doibase 10.1016/j.physletb.2014.09.062} {\bibfield
  {journal} {\bibinfo  {journal} {Phys. Lett. B}\ }\textbf {\bibinfo {volume}
  {738}},\ \bibinfo {pages} {334} (\bibinfo {year} {2014})},\ \Eprint
  {http://arxiv.org/abs/1406.0664} {arXiv:1406.0664 [hep-ph]} \BibitemShut
  {NoStop}%
\bibitem [{\citenamefont {Wang}\ and\ \citenamefont
  {Shen}(2016)}]{Wang:2015ndk}%
  \BibitemOpen
  \bibfield  {author} {\bibinfo {author} {\bibfnamefont {Y.-M.}\ \bibnamefont
  {Wang}}\ and\ \bibinfo {author} {\bibfnamefont {Y.-L.}\ \bibnamefont
  {Shen}},\ }\href {\doibase 10.1007/JHEP02(2016)179} {\bibfield  {journal}
  {\bibinfo  {journal} {JHEP}\ }\textbf {\bibinfo {volume} {02}},\ \bibinfo
  {pages} {179} (\bibinfo {year} {2016})},\ \Eprint
  {http://arxiv.org/abs/1511.09036} {arXiv:1511.09036 [hep-ph]} \BibitemShut
  {NoStop}%
\bibitem [{\citenamefont {Boos}\ \emph
  {et~al.}(2006{\natexlab{a}})\citenamefont {Boos}, \citenamefont {Feldmann},
  \citenamefont {Mannel},\ and\ \citenamefont {Pecjak}}]{Boos:2005by}%
  \BibitemOpen
  \bibfield  {author} {\bibinfo {author} {\bibfnamefont {H.}~\bibnamefont
  {Boos}}, \bibinfo {author} {\bibfnamefont {T.}~\bibnamefont {Feldmann}},
  \bibinfo {author} {\bibfnamefont {T.}~\bibnamefont {Mannel}}, \ and\ \bibinfo
  {author} {\bibfnamefont {B.~D.}\ \bibnamefont {Pecjak}},\ }\href {\doibase
  10.1103/PhysRevD.73.036003} {\bibfield  {journal} {\bibinfo  {journal} {Phys.
  Rev. D}\ }\textbf {\bibinfo {volume} {73}},\ \bibinfo {pages} {036003}
  (\bibinfo {year} {2006}{\natexlab{a}})},\ \Eprint
  {http://arxiv.org/abs/hep-ph/0504005} {arXiv:hep-ph/0504005} \BibitemShut
  {NoStop}%
\bibitem [{\citenamefont {Boos}\ \emph
  {et~al.}(2006{\natexlab{b}})\citenamefont {Boos}, \citenamefont {Feldmann},
  \citenamefont {Mannel},\ and\ \citenamefont {Pecjak}}]{Boos:2005qx}%
  \BibitemOpen
  \bibfield  {author} {\bibinfo {author} {\bibfnamefont {H.}~\bibnamefont
  {Boos}}, \bibinfo {author} {\bibfnamefont {T.}~\bibnamefont {Feldmann}},
  \bibinfo {author} {\bibfnamefont {T.}~\bibnamefont {Mannel}}, \ and\ \bibinfo
  {author} {\bibfnamefont {B.~D.}\ \bibnamefont {Pecjak}},\ }\href {\doibase
  10.1088/1126-6708/2006/05/056} {\bibfield  {journal} {\bibinfo  {journal}
  {JHEP}\ }\textbf {\bibinfo {volume} {05}},\ \bibinfo {pages} {056} (\bibinfo
  {year} {2006}{\natexlab{b}})},\ \Eprint {http://arxiv.org/abs/hep-ph/0512157}
  {arXiv:hep-ph/0512157} \BibitemShut {NoStop}%
\bibitem [{\citenamefont {Wang}\ \emph {et~al.}(2017)\citenamefont {Wang},
  \citenamefont {Wei}, \citenamefont {Shen},\ and\ \citenamefont
  {L\"u}}]{Wang:2017jow}%
  \BibitemOpen
  \bibfield  {author} {\bibinfo {author} {\bibfnamefont {Y.-M.}\ \bibnamefont
  {Wang}}, \bibinfo {author} {\bibfnamefont {Y.-B.}\ \bibnamefont {Wei}},
  \bibinfo {author} {\bibfnamefont {Y.-L.}\ \bibnamefont {Shen}}, \ and\
  \bibinfo {author} {\bibfnamefont {C.-D.}\ \bibnamefont {L\"u}},\ }\href
  {\doibase 10.1007/JHEP06(2017)062} {\bibfield  {journal} {\bibinfo  {journal}
  {JHEP}\ }\textbf {\bibinfo {volume} {06}},\ \bibinfo {pages} {062} (\bibinfo
  {year} {2017})},\ \Eprint {http://arxiv.org/abs/1701.06810} {arXiv:1701.06810
  [hep-ph]} \BibitemShut {NoStop}%
\bibitem [{\citenamefont {Gao}\ \emph {et~al.}(2022)\citenamefont {Gao},
  \citenamefont {Huber}, \citenamefont {Ji}, \citenamefont {Wang},
  \citenamefont {Wang},\ and\ \citenamefont {Wei}}]{Gao:2021sav}%
  \BibitemOpen
  \bibfield  {author} {\bibinfo {author} {\bibfnamefont {J.}~\bibnamefont
  {Gao}}, \bibinfo {author} {\bibfnamefont {T.}~\bibnamefont {Huber}}, \bibinfo
  {author} {\bibfnamefont {Y.}~\bibnamefont {Ji}}, \bibinfo {author}
  {\bibfnamefont {C.}~\bibnamefont {Wang}}, \bibinfo {author} {\bibfnamefont
  {Y.-M.}\ \bibnamefont {Wang}}, \ and\ \bibinfo {author} {\bibfnamefont
  {Y.-B.}\ \bibnamefont {Wei}},\ }\href {\doibase 10.1007/JHEP05(2022)024}
  {\bibfield  {journal} {\bibinfo  {journal} {JHEP}\ }\textbf {\bibinfo
  {volume} {05}},\ \bibinfo {pages} {024} (\bibinfo {year} {2022})},\ \Eprint
  {http://arxiv.org/abs/2112.12674} {arXiv:2112.12674 [hep-ph]} \BibitemShut
  {NoStop}%
\bibitem [{\citenamefont {Beneke}\ \emph {et~al.}(2000)\citenamefont {Beneke},
  \citenamefont {Buchalla}, \citenamefont {Neubert},\ and\ \citenamefont
  {Sachrajda}}]{Beneke:2000ry}%
  \BibitemOpen
  \bibfield  {author} {\bibinfo {author} {\bibfnamefont {M.}~\bibnamefont
  {Beneke}}, \bibinfo {author} {\bibfnamefont {G.}~\bibnamefont {Buchalla}},
  \bibinfo {author} {\bibfnamefont {M.}~\bibnamefont {Neubert}}, \ and\
  \bibinfo {author} {\bibfnamefont {C.~T.}\ \bibnamefont {Sachrajda}},\ }\href
  {\doibase 10.1016/S0550-3213(00)00559-9} {\bibfield  {journal} {\bibinfo
  {journal} {Nucl. Phys. B}\ }\textbf {\bibinfo {volume} {591}},\ \bibinfo
  {pages} {313} (\bibinfo {year} {2000})},\ \Eprint
  {http://arxiv.org/abs/hep-ph/0006124} {arXiv:hep-ph/0006124} \BibitemShut
  {NoStop}%
\bibitem [{\citenamefont {Huber}\ \emph {et~al.}(2016)\citenamefont {Huber},
  \citenamefont {Kr\"ankl},\ and\ \citenamefont {Li}}]{Huber:2016xod}%
  \BibitemOpen
  \bibfield  {author} {\bibinfo {author} {\bibfnamefont {T.}~\bibnamefont
  {Huber}}, \bibinfo {author} {\bibfnamefont {S.}~\bibnamefont {Kr\"ankl}}, \
  and\ \bibinfo {author} {\bibfnamefont {X.-Q.}\ \bibnamefont {Li}},\ }\href
  {\doibase 10.1007/JHEP09(2016)112} {\bibfield  {journal} {\bibinfo  {journal}
  {JHEP}\ }\textbf {\bibinfo {volume} {09}},\ \bibinfo {pages} {112} (\bibinfo
  {year} {2016})},\ \Eprint {http://arxiv.org/abs/1606.02888} {arXiv:1606.02888
  [hep-ph]} \BibitemShut {NoStop}%
\bibitem [{\citenamefont {Benzke}\ \emph {et~al.}(2010)\citenamefont {Benzke},
  \citenamefont {Lee}, \citenamefont {Neubert},\ and\ \citenamefont
  {Paz}}]{Benzke:2010js}%
  \BibitemOpen
  \bibfield  {author} {\bibinfo {author} {\bibfnamefont {M.}~\bibnamefont
  {Benzke}}, \bibinfo {author} {\bibfnamefont {S.~J.}\ \bibnamefont {Lee}},
  \bibinfo {author} {\bibfnamefont {M.}~\bibnamefont {Neubert}}, \ and\
  \bibinfo {author} {\bibfnamefont {G.}~\bibnamefont {Paz}},\ }\href {\doibase
  10.1007/JHEP08(2010)099} {\bibfield  {journal} {\bibinfo  {journal} {JHEP}\
  }\textbf {\bibinfo {volume} {08}},\ \bibinfo {pages} {099} (\bibinfo {year}
  {2010})},\ \Eprint {http://arxiv.org/abs/1003.5012} {arXiv:1003.5012
  [hep-ph]} \BibitemShut {NoStop}%
\bibitem [{\citenamefont {Kawamura}\ \emph {et~al.}(2001)\citenamefont
  {Kawamura}, \citenamefont {Kodaira}, \citenamefont {Qiao},\ and\
  \citenamefont {Tanaka}}]{Kawamura:2001jm}%
  \BibitemOpen
  \bibfield  {author} {\bibinfo {author} {\bibfnamefont {H.}~\bibnamefont
  {Kawamura}}, \bibinfo {author} {\bibfnamefont {J.}~\bibnamefont {Kodaira}},
  \bibinfo {author} {\bibfnamefont {C.-F.}\ \bibnamefont {Qiao}}, \ and\
  \bibinfo {author} {\bibfnamefont {K.}~\bibnamefont {Tanaka}},\ }\href
  {\doibase 10.1016/S0370-2693(01)01299-0} {\bibfield  {journal} {\bibinfo
  {journal} {Phys. Lett. B}\ }\textbf {\bibinfo {volume} {523}},\ \bibinfo
  {pages} {111} (\bibinfo {year} {2001})},\ \bibinfo {note} {[Erratum:
  Phys.Lett.B 536, 344--344 (2002)]},\ \Eprint
  {http://arxiv.org/abs/hep-ph/0109181} {arXiv:hep-ph/0109181} \BibitemShut
  {NoStop}%
\bibitem [{\citenamefont {Braun}\ \emph {et~al.}(2017)\citenamefont {Braun},
  \citenamefont {Ji},\ and\ \citenamefont {Manashov}}]{Braun:2017liq}%
  \BibitemOpen
  \bibfield  {author} {\bibinfo {author} {\bibfnamefont {V.~M.}\ \bibnamefont
  {Braun}}, \bibinfo {author} {\bibfnamefont {Y.}~\bibnamefont {Ji}}, \ and\
  \bibinfo {author} {\bibfnamefont {A.~N.}\ \bibnamefont {Manashov}},\ }\href
  {\doibase 10.1007/JHEP05(2017)022} {\bibfield  {journal} {\bibinfo  {journal}
  {JHEP}\ }\textbf {\bibinfo {volume} {05}},\ \bibinfo {pages} {022} (\bibinfo
  {year} {2017})},\ \Eprint {http://arxiv.org/abs/1703.02446} {arXiv:1703.02446
  [hep-ph]} \BibitemShut {NoStop}%
\bibitem [{\citenamefont {Grinstein}\ \emph {et~al.}(1990)\citenamefont
  {Grinstein}, \citenamefont {Springer},\ and\ \citenamefont
  {Wise}}]{Grinstein:1990tj}%
  \BibitemOpen
  \bibfield  {author} {\bibinfo {author} {\bibfnamefont {B.}~\bibnamefont
  {Grinstein}}, \bibinfo {author} {\bibfnamefont {R.~P.}\ \bibnamefont
  {Springer}}, \ and\ \bibinfo {author} {\bibfnamefont {M.~B.}\ \bibnamefont
  {Wise}},\ }\href {\doibase 10.1016/0550-3213(90)90350-M} {\bibfield
  {journal} {\bibinfo  {journal} {Nucl. Phys. B}\ }\textbf {\bibinfo {volume}
  {339}},\ \bibinfo {pages} {269} (\bibinfo {year} {1990})}\BibitemShut
  {NoStop}%
\bibitem [{\citenamefont {Politzer}(1980)}]{Politzer:1980me}%
  \BibitemOpen
  \bibfield  {author} {\bibinfo {author} {\bibfnamefont {H.~D.}\ \bibnamefont
  {Politzer}},\ }\href {\doibase 10.1016/0550-3213(80)90172-8} {\bibfield
  {journal} {\bibinfo  {journal} {Nucl. Phys. B}\ }\textbf {\bibinfo {volume}
  {172}},\ \bibinfo {pages} {349} (\bibinfo {year} {1980})}\BibitemShut
  {NoStop}%
\bibitem [{\citenamefont {Chetyrkin}\ \emph {et~al.}(1997)\citenamefont
  {Chetyrkin}, \citenamefont {Misiak},\ and\ \citenamefont
  {Munz}}]{Chetyrkin:1996vx}%
  \BibitemOpen
  \bibfield  {author} {\bibinfo {author} {\bibfnamefont {K.~G.}\ \bibnamefont
  {Chetyrkin}}, \bibinfo {author} {\bibfnamefont {M.}~\bibnamefont {Misiak}}, \
  and\ \bibinfo {author} {\bibfnamefont {M.}~\bibnamefont {Munz}},\ }\href
  {\doibase 10.1016/S0370-2693(97)00324-9} {\bibfield  {journal} {\bibinfo
  {journal} {Phys. Lett. B}\ }\textbf {\bibinfo {volume} {400}},\ \bibinfo
  {pages} {206} (\bibinfo {year} {1997})},\ \bibinfo {note} {[Erratum:
  Phys.Lett.B 425, 414 (1998)]},\ \Eprint {http://arxiv.org/abs/hep-ph/9612313}
  {arXiv:hep-ph/9612313} \BibitemShut {NoStop}%
\bibitem [{\citenamefont {Eichten}\ and\ \citenamefont
  {Hill}(1990)}]{Eichten:1989zv}%
  \BibitemOpen
  \bibfield  {author} {\bibinfo {author} {\bibfnamefont {E.}~\bibnamefont
  {Eichten}}\ and\ \bibinfo {author} {\bibfnamefont {B.~R.}\ \bibnamefont
  {Hill}},\ }\href {\doibase 10.1016/0370-2693(90)92049-O} {\bibfield
  {journal} {\bibinfo  {journal} {Phys. Lett. B}\ }\textbf {\bibinfo {volume}
  {234}},\ \bibinfo {pages} {511} (\bibinfo {year} {1990})}\BibitemShut
  {NoStop}%
\bibitem [{\citenamefont {Lunghi}\ \emph {et~al.}(2003)\citenamefont {Lunghi},
  \citenamefont {Pirjol},\ and\ \citenamefont {Wyler}}]{Lunghi:2002ju}%
  \BibitemOpen
  \bibfield  {author} {\bibinfo {author} {\bibfnamefont {E.}~\bibnamefont
  {Lunghi}}, \bibinfo {author} {\bibfnamefont {D.}~\bibnamefont {Pirjol}}, \
  and\ \bibinfo {author} {\bibfnamefont {D.}~\bibnamefont {Wyler}},\ }\href
  {\doibase 10.1016/S0550-3213(02)01032-5} {\bibfield  {journal} {\bibinfo
  {journal} {Nucl. Phys. B}\ }\textbf {\bibinfo {volume} {649}},\ \bibinfo
  {pages} {349} (\bibinfo {year} {2003})},\ \Eprint
  {http://arxiv.org/abs/hep-ph/0210091} {arXiv:hep-ph/0210091} \BibitemShut
  {NoStop}%
\bibitem [{\citenamefont {Bosch}\ \emph {et~al.}(2003)\citenamefont {Bosch},
  \citenamefont {Hill}, \citenamefont {Lange},\ and\ \citenamefont
  {Neubert}}]{Bosch:2003fc}%
  \BibitemOpen
  \bibfield  {author} {\bibinfo {author} {\bibfnamefont {S.~W.}\ \bibnamefont
  {Bosch}}, \bibinfo {author} {\bibfnamefont {R.~J.}\ \bibnamefont {Hill}},
  \bibinfo {author} {\bibfnamefont {B.~O.}\ \bibnamefont {Lange}}, \ and\
  \bibinfo {author} {\bibfnamefont {M.}~\bibnamefont {Neubert}},\ }\href
  {\doibase 10.1103/PhysRevD.67.094014} {\bibfield  {journal} {\bibinfo
  {journal} {Phys. Rev. D}\ }\textbf {\bibinfo {volume} {67}},\ \bibinfo
  {pages} {094014} (\bibinfo {year} {2003})},\ \Eprint
  {http://arxiv.org/abs/hep-ph/0301123} {arXiv:hep-ph/0301123} \BibitemShut
  {NoStop}%
\bibitem [{\citenamefont {Falk}\ \emph {et~al.}(1990)\citenamefont {Falk},
  \citenamefont {Georgi}, \citenamefont {Grinstein},\ and\ \citenamefont
  {Wise}}]{Falk:1990yz}%
  \BibitemOpen
  \bibfield  {author} {\bibinfo {author} {\bibfnamefont {A.~F.}\ \bibnamefont
  {Falk}}, \bibinfo {author} {\bibfnamefont {H.}~\bibnamefont {Georgi}},
  \bibinfo {author} {\bibfnamefont {B.}~\bibnamefont {Grinstein}}, \ and\
  \bibinfo {author} {\bibfnamefont {M.~B.}\ \bibnamefont {Wise}},\ }\href
  {\doibase 10.1016/0550-3213(90)90591-Z} {\bibfield  {journal} {\bibinfo
  {journal} {Nucl. Phys. B}\ }\textbf {\bibinfo {volume} {343}},\ \bibinfo
  {pages} {1} (\bibinfo {year} {1990})}\BibitemShut {NoStop}%
\bibitem [{\citenamefont {Beneke}\ \emph {et~al.}(2020)\citenamefont {Beneke},
  \citenamefont {Bobeth},\ and\ \citenamefont {Wang}}]{Beneke:2020fot}%
  \BibitemOpen
  \bibfield  {author} {\bibinfo {author} {\bibfnamefont {M.}~\bibnamefont
  {Beneke}}, \bibinfo {author} {\bibfnamefont {C.}~\bibnamefont {Bobeth}}, \
  and\ \bibinfo {author} {\bibfnamefont {Y.-M.}\ \bibnamefont {Wang}},\ }\href
  {\doibase 10.1007/JHEP12(2020)148} {\bibfield  {journal} {\bibinfo  {journal}
  {JHEP}\ }\textbf {\bibinfo {volume} {12}},\ \bibinfo {pages} {148} (\bibinfo
  {year} {2020})},\ \Eprint {http://arxiv.org/abs/2008.12494} {arXiv:2008.12494
  [hep-ph]} \BibitemShut {NoStop}%
\bibitem [{\citenamefont {Wang}\ and\ \citenamefont
  {Shen}(2018)}]{Wang:2018wfj}%
  \BibitemOpen
  \bibfield  {author} {\bibinfo {author} {\bibfnamefont {Y.-M.}\ \bibnamefont
  {Wang}}\ and\ \bibinfo {author} {\bibfnamefont {Y.-L.}\ \bibnamefont
  {Shen}},\ }\href {\doibase 10.1007/JHEP05(2018)184} {\bibfield  {journal}
  {\bibinfo  {journal} {JHEP}\ }\textbf {\bibinfo {volume} {05}},\ \bibinfo
  {pages} {184} (\bibinfo {year} {2018})},\ \Eprint
  {http://arxiv.org/abs/1803.06667} {arXiv:1803.06667 [hep-ph]} \BibitemShut
  {NoStop}%
\bibitem [{\citenamefont {Braun}\ and\ \citenamefont
  {Khodjamirian}(2013)}]{Braun:2012kp}%
  \BibitemOpen
  \bibfield  {author} {\bibinfo {author} {\bibfnamefont {V.~M.}\ \bibnamefont
  {Braun}}\ and\ \bibinfo {author} {\bibfnamefont {A.}~\bibnamefont
  {Khodjamirian}},\ }\href {\doibase 10.1016/j.physletb.2012.11.047} {\bibfield
   {journal} {\bibinfo  {journal} {Phys. Lett. B}\ }\textbf {\bibinfo {volume}
  {718}},\ \bibinfo {pages} {1014} (\bibinfo {year} {2013})},\ \Eprint
  {http://arxiv.org/abs/1210.4453} {arXiv:1210.4453 [hep-ph]} \BibitemShut
  {NoStop}%
\bibitem [{\citenamefont {Wang}(2016)}]{Wang:2016qii}%
  \BibitemOpen
  \bibfield  {author} {\bibinfo {author} {\bibfnamefont {Y.-M.}\ \bibnamefont
  {Wang}},\ }\href {\doibase 10.1007/JHEP09(2016)159} {\bibfield  {journal}
  {\bibinfo  {journal} {JHEP}\ }\textbf {\bibinfo {volume} {09}},\ \bibinfo
  {pages} {159} (\bibinfo {year} {2016})},\ \Eprint
  {http://arxiv.org/abs/1606.03080} {arXiv:1606.03080 [hep-ph]} \BibitemShut
  {NoStop}%
\bibitem [{\citenamefont {Grozin}\ and\ \citenamefont
  {Neubert}(1997)}]{Grozin:1996pq}%
  \BibitemOpen
  \bibfield  {author} {\bibinfo {author} {\bibfnamefont {A.~G.}\ \bibnamefont
  {Grozin}}\ and\ \bibinfo {author} {\bibfnamefont {M.}~\bibnamefont
  {Neubert}},\ }\href {\doibase 10.1103/PhysRevD.55.272} {\bibfield  {journal}
  {\bibinfo  {journal} {Phys. Rev. D}\ }\textbf {\bibinfo {volume} {55}},\
  \bibinfo {pages} {272} (\bibinfo {year} {1997})},\ \Eprint
  {http://arxiv.org/abs/hep-ph/9607366} {arXiv:hep-ph/9607366} \BibitemShut
  {NoStop}%
\bibitem [{\citenamefont {Braun}\ \emph {et~al.}(2004)\citenamefont {Braun},
  \citenamefont {Ivanov},\ and\ \citenamefont {Korchemsky}}]{Braun:2003wx}%
  \BibitemOpen
  \bibfield  {author} {\bibinfo {author} {\bibfnamefont {V.~M.}\ \bibnamefont
  {Braun}}, \bibinfo {author} {\bibfnamefont {D.~Y.}\ \bibnamefont {Ivanov}}, \
  and\ \bibinfo {author} {\bibfnamefont {G.~P.}\ \bibnamefont {Korchemsky}},\
  }\href {\doibase 10.1103/PhysRevD.69.034014} {\bibfield  {journal} {\bibinfo
  {journal} {Phys. Rev. D}\ }\textbf {\bibinfo {volume} {69}},\ \bibinfo
  {pages} {034014} (\bibinfo {year} {2004})},\ \Eprint
  {http://arxiv.org/abs/hep-ph/0309330} {arXiv:hep-ph/0309330} \BibitemShut
  {NoStop}%
\bibitem [{\citenamefont {Khodjamirian}\ \emph {et~al.}(2007)\citenamefont
  {Khodjamirian}, \citenamefont {Mannel},\ and\ \citenamefont
  {Offen}}]{Khodjamirian:2006st}%
  \BibitemOpen
  \bibfield  {author} {\bibinfo {author} {\bibfnamefont {A.}~\bibnamefont
  {Khodjamirian}}, \bibinfo {author} {\bibfnamefont {T.}~\bibnamefont
  {Mannel}}, \ and\ \bibinfo {author} {\bibfnamefont {N.}~\bibnamefont
  {Offen}},\ }\href {\doibase 10.1103/PhysRevD.75.054013} {\bibfield  {journal}
  {\bibinfo  {journal} {Phys. Rev. D}\ }\textbf {\bibinfo {volume} {75}},\
  \bibinfo {pages} {054013} (\bibinfo {year} {2007})},\ \Eprint
  {http://arxiv.org/abs/hep-ph/0611193} {arXiv:hep-ph/0611193} \BibitemShut
  {NoStop}%
\bibitem [{\citenamefont {L\"u}\ \emph {et~al.}(2019)\citenamefont {L\"u},
  \citenamefont {Shen}, \citenamefont {Wang},\ and\ \citenamefont
  {Wei}}]{Lu:2018cfc}%
  \BibitemOpen
  \bibfield  {author} {\bibinfo {author} {\bibfnamefont {C.-D.}\ \bibnamefont
  {L\"u}}, \bibinfo {author} {\bibfnamefont {Y.-L.}\ \bibnamefont {Shen}},
  \bibinfo {author} {\bibfnamefont {Y.-M.}\ \bibnamefont {Wang}}, \ and\
  \bibinfo {author} {\bibfnamefont {Y.-B.}\ \bibnamefont {Wei}},\ }\href
  {\doibase 10.1007/JHEP01(2019)024} {\bibfield  {journal} {\bibinfo  {journal}
  {JHEP}\ }\textbf {\bibinfo {volume} {01}},\ \bibinfo {pages} {024} (\bibinfo
  {year} {2019})},\ \Eprint {http://arxiv.org/abs/1810.00819} {arXiv:1810.00819
  [hep-ph]} \BibitemShut {NoStop}%
\bibitem [{\citenamefont {Khodjamirian}\ \emph {et~al.}(2020)\citenamefont
  {Khodjamirian}, \citenamefont {Mandal},\ and\ \citenamefont
  {Mannel}}]{Khodjamirian:2020hob}%
  \BibitemOpen
  \bibfield  {author} {\bibinfo {author} {\bibfnamefont {A.}~\bibnamefont
  {Khodjamirian}}, \bibinfo {author} {\bibfnamefont {R.}~\bibnamefont
  {Mandal}}, \ and\ \bibinfo {author} {\bibfnamefont {T.}~\bibnamefont
  {Mannel}},\ }\href {\doibase 10.1007/JHEP10(2020)043} {\bibfield  {journal}
  {\bibinfo  {journal} {JHEP}\ }\textbf {\bibinfo {volume} {10}},\ \bibinfo
  {pages} {043} (\bibinfo {year} {2020})},\ \Eprint
  {http://arxiv.org/abs/2008.03935} {arXiv:2008.03935 [hep-ph]} \BibitemShut
  {NoStop}%
\bibitem [{\citenamefont {Beneke}\ \emph
  {et~al.}(2018{\natexlab{a}})\citenamefont {Beneke}, \citenamefont {Braun},
  \citenamefont {Ji},\ and\ \citenamefont {Wei}}]{Beneke:2018wjp}%
  \BibitemOpen
  \bibfield  {author} {\bibinfo {author} {\bibfnamefont {M.}~\bibnamefont
  {Beneke}}, \bibinfo {author} {\bibfnamefont {V.~M.}\ \bibnamefont {Braun}},
  \bibinfo {author} {\bibfnamefont {Y.}~\bibnamefont {Ji}}, \ and\ \bibinfo
  {author} {\bibfnamefont {Y.-B.}\ \bibnamefont {Wei}},\ }\href {\doibase
  10.1007/JHEP07(2018)154} {\bibfield  {journal} {\bibinfo  {journal} {JHEP}\
  }\textbf {\bibinfo {volume} {07}},\ \bibinfo {pages} {154} (\bibinfo {year}
  {2018}{\natexlab{a}})},\ \Eprint {http://arxiv.org/abs/1804.04962}
  {arXiv:1804.04962 [hep-ph]} \BibitemShut {NoStop}%
\bibitem [{\citenamefont {Beneke}\ \emph
  {et~al.}(2018{\natexlab{b}})\citenamefont {Beneke}, \citenamefont {Bobeth},\
  and\ \citenamefont {Szafron}}]{Beneke:2017vpq}%
  \BibitemOpen
  \bibfield  {author} {\bibinfo {author} {\bibfnamefont {M.}~\bibnamefont
  {Beneke}}, \bibinfo {author} {\bibfnamefont {C.}~\bibnamefont {Bobeth}}, \
  and\ \bibinfo {author} {\bibfnamefont {R.}~\bibnamefont {Szafron}},\ }\href
  {\doibase 10.1103/PhysRevLett.120.011801} {\bibfield  {journal} {\bibinfo
  {journal} {Phys. Rev. Lett.}\ }\textbf {\bibinfo {volume} {120}},\ \bibinfo
  {pages} {011801} (\bibinfo {year} {2018}{\natexlab{b}})},\ \Eprint
  {http://arxiv.org/abs/1708.09152} {arXiv:1708.09152 [hep-ph]} \BibitemShut
  {NoStop}%
\bibitem [{\citenamefont {Cerri}\ \emph {et~al.}(2019)\citenamefont {Cerri}
  \emph {et~al.}}]{Cerri:2018ypt}%
  \BibitemOpen
  \bibfield  {author} {\bibinfo {author} {\bibfnamefont {A.}~\bibnamefont
  {Cerri}} \emph {et~al.},\ }\href {\doibase 10.23731/CYRM-2019-007.867}
  {\bibfield  {journal} {\bibinfo  {journal} {CERN Yellow Rep. Monogr.}\
  }\textbf {\bibinfo {volume} {7}},\ \bibinfo {pages} {867} (\bibinfo {year}
  {2019})},\ \Eprint {http://arxiv.org/abs/1812.07638} {arXiv:1812.07638
  [hep-ph]} \BibitemShut {NoStop}%
\bibitem [{\citenamefont {Ciuchini}\ \emph
  {et~al.}(2021{\natexlab{b}})\citenamefont {Ciuchini}, \citenamefont {Fedele},
  \citenamefont {Franco}, \citenamefont {Paul}, \citenamefont {Silvestrini},\
  and\ \citenamefont {Valli}}]{Ciuchini:2021smi}%
  \BibitemOpen
  \bibfield  {author} {\bibinfo {author} {\bibfnamefont {M.}~\bibnamefont
  {Ciuchini}}, \bibinfo {author} {\bibfnamefont {M.}~\bibnamefont {Fedele}},
  \bibinfo {author} {\bibfnamefont {E.}~\bibnamefont {Franco}}, \bibinfo
  {author} {\bibfnamefont {A.}~\bibnamefont {Paul}}, \bibinfo {author}
  {\bibfnamefont {L.}~\bibnamefont {Silvestrini}}, \ and\ \bibinfo {author}
  {\bibfnamefont {M.}~\bibnamefont {Valli}},\ }\href@noop {} {\  (\bibinfo
  {year} {2021}{\natexlab{b}})},\ \Eprint {http://arxiv.org/abs/2110.10126}
  {arXiv:2110.10126 [hep-ph]} \BibitemShut {NoStop}%
\bibitem [{\citenamefont {Altmannshofer}\ and\ \citenamefont
  {Archilli}(2022)}]{Altmannshofer:2022hfs}%
  \BibitemOpen
  \bibfield  {author} {\bibinfo {author} {\bibfnamefont {W.}~\bibnamefont
  {Altmannshofer}}\ and\ \bibinfo {author} {\bibfnamefont {F.}~\bibnamefont
  {Archilli}},\ }in\ \href@noop {} {\emph {\bibinfo {booktitle} {{2022 Snowmass
  Summer Study}}}}\ (\bibinfo {year} {2022})\ \Eprint
  {http://arxiv.org/abs/2206.11331} {arXiv:2206.11331 [hep-ph]} \BibitemShut
  {NoStop}%
\bibitem [{\citenamefont {Altmannshofer}\ \emph {et~al.}(2019)\citenamefont
  {Altmannshofer} \emph {et~al.}}]{Belle-II:2018jsg}%
  \BibitemOpen
  \bibfield  {author} {\bibinfo {author} {\bibfnamefont {W.}~\bibnamefont
  {Altmannshofer}} \emph {et~al.} (\bibinfo {collaboration} {Belle-II}),\
  }\href {\doibase 10.1093/ptep/ptz106} {\bibfield  {journal} {\bibinfo
  {journal} {PTEP}\ }\textbf {\bibinfo {volume} {2019}},\ \bibinfo {pages}
  {123C01} (\bibinfo {year} {2019})},\ \bibinfo {note} {[Erratum: PTEP 2020,
  029201 (2020)]},\ \Eprint {http://arxiv.org/abs/1808.10567} {arXiv:1808.10567
  [hep-ex]} \BibitemShut {NoStop}%
\end{thebibliography}%

\end{document}